\documentclass[12pt]{article}
\usepackage{latexsym}
\usepackage{amssymb}
\usepackage{amsmath}
\usepackage{graphicx}
\usepackage{enumitem}
\usepackage{bbold}
\usepackage{slashed}

\newcommand\eeq{\end{equation}} 
\newcommand\beq{\begin{equation}}

\newcommand\mathC{\mkern1mu\raise2.2pt\hbox{$\scriptscriptstyle|$}
        {\mkern-7mu\rm C}}              
\newcommand{\mathR}{{\rm I\! R}}         

\def\be{\begin{equation}}
\def\ee{\end{equation}}
\def\bear{\begin{eqnarray}}
\def\eear{\end{eqnarray}}

\newcommand\bra[1]{{\langle {#1}|}}
\newcommand\ket[1]{{|{#1}\rangle}}

\def\bra{\langle}
\def\ket{\rangle}

\newcommand{\sm}[1]{\mbox{\scriptsize #1}}

\renewcommand{\@}[1]{\sqrt{#1}}
\renewcommand{\le}[1]{\label{#1}\end{eqnarray}}
\newcommand{\bea}{\begin{eqnarray}}
\newcommand{\eea}{\end{eqnarray}}

\def\ffract#1#2{\raise .35 em\hbox{$\scriptstyle#1$}\kern-.25em/
\kern-.2em\lower .22 em \hbox{$\scriptstyle#2$}}

\begin{document}

\pagestyle{empty}

\centerline{{\Large \bf On Dualities  and Equivalences Between Physical Theories}}
\vskip 1.5truecm

\begin{center}
{\large  
Jeremy Butterfield: Monday 28 May 2018}\\
\date{\today}
\vskip 1truecm
{\it Trinity College, Cambridge, CB2 1TQ, United Kingdom}\\
\vskip .5truecm
{\tt jb56@cam.ac.uk}\\

\vskip 1truecm
Dedicated to Graeme Segal. Forthcoming (abridged) in {\em Philosophy Beyond Spacetime} (OUP), ed.s N. Huggett, B. Le Bihan and C. W\"{u}thrich.

\vskip 1truecm
Motto: `I defy anyone to avoid getting confused by active vs. passive  transformations' (Graeme Segal, in conversation about Einstein's hole argument, 2006).
\end{center}

\vskip 3truecm

\begin{center}
\textbf{\large \bf Abstract}
\end{center}
The main aim of this paper is to make a remark about the relation between (i) dualities between theories, as `duality' is understood in physics and (ii) equivalence of theories, as `equivalence' is understood in logic and philosophy. The remark is that in physics, two theories can be dual, and accordingly get called `the same theory', though we interpret them as disagreeing---so that they are certainly {\em not} equivalent,  as `equivalent' is normally understood.  So the remark is simple: but, I shall argue, worth stressing---since often neglected.

My argument for this is based on the account of duality developed by De Haro: which is illustrated here with several examples, from both elementary physics and string theory. Thus I argue that in some examples, including in string theory, two dual theories disagree in their claims about the world. 

I also spell out how this remark implies a limitation of proposals (both traditional and recent) to understand theoretical equivalence as either logical equivalence or a weakening of it.

\newpage
\pagestyle{plain}

\tableofcontents

\newpage

\section{Introduction}\label{intro}
\subsection{Prospectus: the Remark and the Implication}\label{prosp}
In recent philosophy of physics, there has been a surge of interest in two topics:\\
\indent \indent (i) how best to understand dualities between physical theories; \\
\indent \indent(ii) how best to individuate physical theories, and so how best to define theories being equivalent. \\
These topics are closely related, since all agree that a duality is a matter of two theories being in some sense `the same'.  
But the main aim of this paper is to stress that in physics, two theories can be dual, and accordingly get called `the same theory', though we interpret them as disagreeing---so that they are certainly {\em not} equivalent,  as `equivalent' is normally understood, both in everyday language and  in  philosophy. 

I will establish this point---I shall say {\em Remark}---by: (a) reporting the account of duality in physics, developed by De Haro (2016, 2016a), which I endorse and which we (2017) call a `Schema' (Sections \ref{schema} and \ref{subject}); and then (b) supporting the Schema with examples from classical  and quantum physics, some of which illustrate the Remark  (Section \ref{weak;exples}). 

As we shall see in these examples, two dual theories can `disagree'  in either of two ways:  either \\
\indent \indent (Contr): by the theories making contrary assertions about a common subject-matter; (they may well agree on some claims, but they contradict each other over other claims); or \\
\indent  \indent (Diff): by the theories describing different subject-matters (though the descriptions are `isomorphic'  or  `matching'---hence the duality).

In Section \ref{subject}, I will make precise the notion of a subject-matter: (although for most of the paper this precision is not needed, it will be useful in Section \ref{string}). I also agree that (Diff) hardly merits the label `disagreement', since both theories could be true. But `disagree' is just my convenient umbrella term. The main point is the same, for either (Contr) or (Diff): in both situations, one would not say that the dual theories are `equivalent', on any normal understanding.

Establishing the Remark leaves two tasks that will occupy the second half of the paper. First, I spell out an {\em Implication} of the Remark: namely, a limitation of proposals (both traditional and recent) to understand theoretical equivalence either as logical equivalence or as a weakening of it (Section \ref{logic}). In short: two disagreeing dual theories might be formalised so as to be logically equivalent. And this implies that logical equivalence is too weak an explication  of theoretical equivalence---as is, therefore, any of the recently proposed weakenings of logical equivalence.  

Second, I briefly discuss the Remark's application to dualities in string theory (Section \ref{string}).  Of course, this is a vast topic, which I cannot properly address.  But my having established the Remark with examples from elementary (classical  and quantum) physics prompts the question: Do dualities in more advanced physics, in particular in string theory, also illustrate it? 

 Indeed, this question is all the more pressing since (so far as De Haro and I know) the one  duality in advanced physics that has been exhibited in detail as an example of our Schema (viz. bosonization, in De Haro  and Butterfield 2017, Sections 4, 5) is, as it happens, a duality in which the two dual theories do {\em not} disagree in either of the senses (Contr) or (Diff). Rather, they are both about a common subject-matter (viz. a quantum field system in 1+1 dimensions, i.e. with space one-dimensional), about which they say different but consistent things. Bosonization is thus a cousin---a very advanced cousin!---of position-momentum duality in elementary quantum mechanics (cf. Section \ref{Ex2}): in which the position and momentum descriptions do not disagree but say different yet consistent things about a common subject-matter. In any case, Section \ref{string} will argue that Yes, some string-theoretic dualities do illustrate the Remark.\\

Thus the paper has three main stages. The first stage, Sections \ref{schema} to \ref{weak;exples}, establishes the Remark, with elementary examples from physics. Though the examples are elementary, the `tone' is philosophy of physics, rather than of logic. In the second stage,  Section \ref{logic}, philosophy of logic comes to the fore. Indeed, readers interested more in logic than in physics can read Section \ref{logic} without reading the details of Sections \ref{schema} to \ref{weak;exples}. The third stage, Section \ref{string}, returns to philosophy of physics, with examples from string theory: examples which are, unfortunately, not as rigorously established as the elementary examples.\\

As we shall see, the Remark is, in essence, simple---`elementary, my dear Watson'---in two ways. First: it is illustrated by some familiar, indeed traditional, cases, which are examples of our  Schema for duality.  An example of (Contr) will be different (i.e. contrary) specifications of absolute rest in Newtonian mechanics, set in Newtonian spacetime. Examples of  (Diff) will be: a ``reworking'' of position-momentum duality in quantum mechanics; and Kramers-Wannier duality in statistical mechanics.  These will be Examples (1), (2) and (3), in Section \ref{weak;exples}. Second:  the Remark, and also the Implication, turn on a familiar idea: that formal mathematical methods can only ``discern structure'', and so  cannot ``cut finer'' than isomorphism, i.e. cannot distinguish isomorphic copies. The classification of an Example as a case of (Contr) (or of (Diff)) therefore goes beyond formal matters such as isomorphism, and depends on interpretation.\\

This appeal to interpretation prompts two more general, but again familiar, morals of what follows. 

The first moral is about the need for judgment---and the difficulty of distinguishing  `fact' from `convention'. In logic class, we learn to respect the gap between formal logical systems and arguments in natural language; and to admit that regimenting such an argument in a system requires some judgment-calls. Similarly here: aligning physical theories with our formal definition of duality will require some judgments. These judgments will impinge on whether we classify an Example as a case of (Contr); and the same goes for (Diff). I am steadfast that I have classified my Examples correctly, by respecting the intended meanings of the theories' words. In Section \ref{subject}, I will make `intended meaning' more precise (viz. by endorsing the framework of intensional semantics). But I admit to relying on intuition---or perhaps better: stipulation---about e.g. what a proponent of Newtonian mechanics means by their word `absolute rest', or what a proponent of statistical mechanics means by `temperature'. But---as we will see---I think this is the best one can do, thanks to lessons philosophers learnt some sixty years ago, from the troubles that beset positivists' attempts to sharply distinguish, within a scientific theory, `fact' from `convention'---its empirical claims from its conventional elements---once and for all.\footnote{{\label{Putnam62}}{Here it is usual to cite Quine (1953). But I maintain that Quine was unfair to Carnap, and that there are more persuasive rebuttals of making this a sharp distinction; cf. Putnam (1962), Stein (1992).}}

The second moral is broader. It is that, quite apart from the topic of dualities (and in particular, De Haro's and my account of them), a verdict that two theories  are equivalent depends on interpretation. `Theoretical equivalence' is of course a term of art in the post-positivist tradition, that takes `scientific theory' as a unit of analysis and accepts some sort of theory/observation distinction. In principle, you could define it however you wish. But all hands agree that we need some notion of `theories  making the very same claims about the world, i.e. not only observational claims but also theoretical ones'. For one thing, we need some such strengthening of the idea of observational (also known as: empirical) equivalence, in order to articulate and then assess the under-determination of theory by data, namely as a matter of observationally equivalent, but  theoretically inequivalent, theories. Here, the phrase `claims about the world'  obviously invokes interpretation. But interpretation is liable to be vague, and even controversial, not least because of widespread post-Quinean suspicion of meanings. Hence the long tradition of using formal ideas from logic to try and give a precise explication of theoretical equivalence. The best-known proposal takes theories in the traditional way as sets of sentences, and then defines {\em logical equivalence} of theories, exactly as in a logic book. 

But this cannot be right as an explication of theoretical equivalence, for three  reasons: each of which leads us back to interpretation. The first two are well-known; and the third is the main idea of this paper's Implication.\\
\indent \indent (1):  Many philosophers reject the  traditional construal of theories as sets of sentences, usually by advocating the rival  `semantic conception of theories' (a view I will touch on again in Section \ref{implic}). \\
\indent \indent (2): For two  theories to be logically equivalent,  in the precise sense of a logic book, they must be formulated in the very same vocabulary. And this is certainly too restrictive an explication of `making the very same claims about the world', since two theories could do that while using different vocabularies. Just think of classical electromagnetism written in English and French. This second point has prompted the effort, in both logic and philosophy, to give weaker definitions: that is, to define equivalence of theories more weakly than (but as precisely as) the familiar logical equivalence. (I review this effort in Section \ref{implic2}.)\\
\indent \indent (3): As I announced above, my Remark has an Implication that `goes in the opposite direction' from this effort. For the Remark shows that logical equivalence is in some cases too weak an explication of `making the very same claims about the world'. Namely: dual pairs satisfying (Contr) or (Diff), once they are formalized (i.e. stated in a formal language), may well be logically equivalent, and thereby equivalent in any weaker sense---though, thanks to (Contr) or (Diff), they certainly do {\em not} `make the very same claims about the world'.\\
\indent \indent  Here, reasons (2) and (3) obviously form a complementary pair. (2) says that we need to allow for {\em synonymy}, like `electric' in English and `\'{e}lectrique' in French; and (3) says that we need to allow for {\em homonymy}, like `bank' in finance and beside a river. As I said in the first moral above: I am steadfast that, though we need to exercise  judgment, in my Examples the facts about synonymy and homonymy are sufficiently objective to make my Remark and Implication true.\footnote{{\label{Bas1}}Of course, I am not alone in stressing that verdicts of theoretical equivalence depend on interpretation. Coffey (2014) argues for this (partly by citing philosophers disagreeing about putative examples: his Sections 2, 3). I agree with him in spirit; (though not in letter---I disagree with his treatment of examples, and I do not share his post-Quinean anxiety about meanings (pp. 835, 838)). A closer kindred spirit is van Fraassen, especially in his reply (2014) to Halvorson's critique (2012) of the semantic view of theories. Van Fraassen's Section 2 stresses the indispensability of interpretation. In particular, he says: `the identity of a theory [cannot] be defined in terms of the corresponding set of mathematical structures without reference to their representational function' (p. 278). And more specifically, van Fraassen prefigures my reason (3) about homonymy: `if the same diffusion equation is presented to describe gas diffusion, and, elsewhere, temperature distribution over time, would anyone think that one and only one theory was being presented?' (p. 279). He also says the point is familiar: `these sorts of examples had ample discussion in familiar literature' (ibid.). As I say in the main text: I agree---but maintain that nowadays, it is still worth stressing, especially in connection with dualities.}  \\

But though the Remark and Implication are simple, and these two general morals familiar: they are worth stressing, for three reasons. First: simple though they are, most of the recent philosophical literature on dualities and theoretical  equivalence {\em fails} to state them: which can be confusing. As we shall see, the confusion is closely related to confusions that often occur when thinking about active vs. passive  transformations: hence the paper's {\em motto}. Second:  they  provide the mental exercise of navigating the various conflicting usages of such words as `theory', `model', `equivalence', and even `isomorphism'. Third: since the interpretation of  string-theoretic dualities is a central topic of the philosophical literature on dualities, it is worth seeing that the Remark is illustrated by such dualities, not just elementary ones. All the more so, since much of that literature denies the Remark.

\subsection{Relations to other work}\label{others}
To help introduce the paper, I comment on relations to other work. The first comment is that this paper is a companion to De Haro (2016, 2016a): not just because these papers state the Schema for duality (in their first Sections); but also because they emphasise that dual theories can disagree (in my umbrella sense). In De Haro's jargon: dual theories are by definition {\em theoretically equivalent}: so note that for him, `theoretical equivalence' does {\em not} mean `making the same claims about the world'. Yet, he says, dual theories  can make different, i.e. disagreeing, claims about the world---which he dubs their being {\em physically inequivalent}. I agree with De Haro, though I shall keep to my jargon, including my colloquial umbrella word `disagree', and (Contr) and (Diff); (Section \ref{jargon} gives my rationale about jargon).

To see what this paper adds to De Haro's work, note first that De Haro proposes a sufficient condition for one to be justified in interpreting two duals to be physically equivalent (in his sense). This sufficient condition has two conjuncts. The first is that each dual is {\em unextendable}: which means, roughly speaking, that the dual, i.e. the theory, both is a complete description of its intended domain and cannot be extended to a larger domain. The second is that each dual has an {\em internal} (as against {\em external}) interpretation: i.e. an interpretation that does {\em not} proceed by coupling to another theory, often one which describes measurements of the given theory's domain. These conjuncts are linked in that De Haro argues that unextendability implies that one is justified in using an internal interpretation: (justified but not obliged---there can still be external interpretations). So to summarize: De Haro proposes that unextendability justifies (but does not force)  internality; and the two together, applied to the duals of a duality, justify interpreting the duals as physically equivalent: (De Haro 2016, Sections 1.3.1, 1.3.2; 2016a, Sections 1.3.1, 1.3.3; 2017, Section 2.4.1).\footnote{Two remarks. (1):  The last reference speaks of internal vs. external  `viewpoints', rather than `interpretations': so also do Dieks et al. (2015: Section 3.3.2). (2): De Haro does not invoke the framework of intensional semantics, as I will in Section \ref{subject}. But his proposed sufficient condition being, not for physical equivalence {\em simpliciter}, but for one being justified in such an interpretation, reflects his taking physical equivalence to be a matter of interpretation, outstripping formal methods.} 

I do not need more details about this proposal (sympathetic though I am). I need only note that, as De Haro also stresses, it is putative `theories of everything' (TOEs), `toy cosmologies', i.e. theories of an entire possible world, that are, or can best claim to be, unextendable. So, since string theories  are usually presented as TOEs (although perhaps not as `final theories', whatever that might mean!), De Haro goes on to discuss string-theoretic dualities, especially  gauge/gravity duality, exploring the details of physical equivalence and internal interpretations (in his senses). (Section \ref{Ex6} will say a bit more, but his main interpretative discussions are: 2016, Sections 2.1, 2.3; 2016a, Sections 2, 3; 2017, Section 2.)

I can now state what this paper adds to De Haro's results, in three points. First, it is worth seeing that dual theories can disagree (for De Haro: be physically inequivalent), in {\em simple} examples (Section \ref{weak;exples}).  For then the Remark is not hidden in the {\em technicalia} of advanced physics; (nor, I might add, is it threatened by  most string-theoretic dualities being not yet rigorously proven, and thereby questionable).  Second, there is the Remark's Implication: a cautionary moral about the limits of formal methods, independent of advanced physics (Section \ref{logic}). Third, when I do turn to string theory, I will argue that for two dualities, viz.  {\em gauge/gravity duality} (Section \ref{Ex6}) and  {\em T-duality} (Section \ref{Ex7}), the two dual theories disagree---in the straightforward `contradiction' sense of (Contr). To put it more modestly: I will argue that this view of the duals is tenable---as tenable as was Newton's and Clarke's view that two different specifications of absolute rest contradict each other. \\

This brings me to this paper's relations to other recent  literature about dualities. Here, I will  just expand on the first and third of my three reasons, at the end of Section \ref{prosp}, for stressing the Remark. (I will later cite some specific connections: about (a) examples like the harmonic oscillator (Section \ref{symmthy}), and Lagrangian vs. Hamiltonian mechanics (Section \ref{Ex5}); and (b) recent logical work on theoretical equivalence (Section \ref{implic2}).) 

These two reasons are linked. The reason why most of this literature omits the Remark (and so its Implication) is that it concentrates on dualities in string theories: which, as I mentioned, are usually presented as TOEs. Thus several authors hold that dualities  for TOEs (`toy cosmologies') {\em must} be taken as making the same claims about the world: as `agreeing', in my jargon---though often called by them, as by De Haro, `physically equivalent'. Of course, the details vary. Some authors do not  use the phrase `physical equivalence' (or `theoretical equivalence') precisely; and they do not articulate De Haro's idea of an internal interpretation as a way of vetoing two duals being about isomorphic but distinct subject-matters. But there is a common intuitive idea: roughly speaking, that for two dual TOEs, there is literally `no room in the cosmos' for them to have separate but isomorphic subject-matters.  Here are three examples:\\
\indent \indent \indent (a): Rickles says `dual descriptions are not in competition for `physical reality': rather they are to be considered as complementary descriptions of one and the same physical situation' (2011: 55), and `theories are said to be dual when they generate the same physics' (2011: 56). And again in later work: `we can take both descriptions to be describing the same physical situation using different  concepts' (2013: 319); `there is no {\em competition} between dual descriptions, and the incompatibility can be transformed away in a certain sense ... dualities are examples of equivalent descriptions {\em simpliciter}' (2017: 63 and {\em passim}).\\
\indent \indent \indent (b): Matsubara (2013: Section 4.1, p. 477-478)  requires dual theories  to be physically equivalent, viz. by requiring any differences between them not to be matters of empirical interpretation, nor choices of coordinates or gauge (or at least not obviously so); \\
\indent \indent \indent (c): Huggett (2017: 86)  says `the case of dual total theories is clearly one in which the putative differences are `hidden' in a very strong sense ...[so that] at least from a practical, scientific point of view, it makes sense to treat those differences as non-physical ... In other words, long established, well-motivated scientific reasoning should lead us to think that dual total theories represent the same physical situation.'\footnote{{\label{WAS2qualif}}{Agreed: these authors also note that in some dualities, away from string theory or TOEs, the duals have different subject-matters. For example, Matsubara (2013: Section 4.2.1, p. 478) and Huggett (2017: 86) note this for an elementary harmonic oscillator duality---which I will discuss in  Section \ref{symmthy}.}}\\

So to sum up this paper's relation to these authors: I will deny this consensus. For Section \ref{string} will urge that even in string theory, the two dual theories disagree. Or more modestly: this view of the duals is tenable---as tenable as was Newton's and Clarke's view that two different specifications of absolute rest contradict each other.  

I of course agree that even for TOEs,  the literature is not unanimous. A laudable exception to the consensus---a close cousin of my own position---is Read (2016), and Read and M\"{o}ller-Nielsen (2018). The first of these is about  string-theoretic dualities. It gives a nuanced roster of interpretative options, and seems to favour a combination of them (2016: 224-225, 232), rather than saying that duals always agree, i.e. are physically equivalent (his options 2 and 3, explored on pp. 227-230). The second paper relates our topic---do some duals disagree?---to a proposed distinction between two general approaches to symmetries and dualities, called the `interpretational approach', and the `motivational approach' (a distinction proposed by M\"{o}ller-Nielsen (2017)). The broad idea is that `interpretationalists' infer confidently from the mere existence of a symmetry or duality to agreement, i.e. physical equivalence; but `motivationalists' are cautious: they say that only once a common ontology is formulated, should one infer agreement. 

Thus Read and M\"{o}ller-Nielsen are cautious motivationalists. Their position is obviously a cousin of mine---and I dare say, of De Haro's: as is clear from their helpful comparison of various positions (their Section 5.3). But I will not go into detail. For: (i) their definition of duality is different (roughly speaking: weaker) than De Haro's and mine (in our Schema); and (ii) their `interpretational vs. motivational' distinction concerns the  {\em heuristic} functions of duality for guessing a better theory `behind' the two duals---functions which I will mostly set aside. But I will agree that these functions are very important: (cf. my warning {\em Beware}  in Section \ref{Ex1}, its footnote \ref{heuristic}, item (5) in Section \ref{Ex6}, and De Haro (2018)). Indeed, these functions will suggest a diagnosis of {\em why} the false consensus above gelled: i.e. why my Remark has been missed. In short: the diagnosis is that the authors are using the duality to guess a better theory `behind' the duals, rather than (like me) interpreting dual theories  as now formulated.

\section{A Schema for duality}\label{schema}
  In this Section and the next, I review our account of duality (Sections \ref{idea}, \ref{details}),  compare it with the notion of symmetry (Section \ref{symmthy}), and relate it to ideas of interpretation (Section \ref{subject}).  This  account has been developed mainly by  De Haro (2016:~Section 1, 2016a:~Section 1), but also more fully by us together (2017: Sections 2, 3):   (and foreshadowed in De Haro, Teh, and Butterfield (2017: Section 3)).  As a mnemonic, we label this account a {\em Schema}. It will be clear that our Schema is logically weak, so that there are countless examples, including elementary ones: a topic taken up in Section \ref{credo}. 

\subsection{The idea}\label{idea}
De Haro and I say that a duality is an isomorphism of theories.
More precisely: duality is a matter of:\\
\indent (a): the two given theories sharing a {\em common core}; the common core is itself a theory, which we call a {\em bare theory}; (it is usually best to think of the bare theory as uninterpreted, or abstract: though it may be interpreted); and \\
\indent (b): the two given theories being {\em isomorphic models} of this common core: that is, isomorphic as regards the structure of the bare theory. The two dual i.e. isomorphic theories  will be called {\em model triples}, the `triple' referring to the fact that the theory consists of three items: a state-space $\cal S$, a set of quantities $\cal Q$, and a dynamics $\cal D$.\\

Here, a model is a specific realization---one might say `version' or `formulation'---of a theory. That is: it `models' (verb!) another theory, viz. the bare theory: usually in the sense of representation theory (in the mathematical sense of `representation theory'---`representation' being another word with confusingly diverse uses, in mathematics and in philosophy). 

So:  {\em Beware}: the word `model' often has three connotations that are for our Schema {\em misleading}. That is: the word `model', as contrasted with `theory', often connotes: \\
\indent \indent  (i): a specific solution (at a single time: or for all times, i.e.~a possible history) for the physical system concerned, whereas the `theory' encompasses all solutions---and in many cases,  for a whole class of systems; \\
\indent \indent  (ii): an approximation, in particular an approximate solution, whereas the `theory' deals with exact solutions; \\
\indent \indent (iii): an interpretation, and even being part of the physical world (in particular, being empirical, and-or observable); whereas the `theory' stands in need of interpretation, and is of course not part of the physical world.

So NB: {\em our use of `model' denies all three connotations}. 

A standard piece of physics jargon can also be misleading: namely, calling a duality's correspondence between items---the mapping of a state or quantity in one dual theory to the other---a {\em dictionary}.   Philosophers, beware: the `dictionary' need not be faithful to meanings. Items that it pairs can have very different meanings. Indeed: this will be crucial to cases of (Contr) and (Diff), and thus to  the Remark and Implication.\\

A rigorous but advanced example of this Schema  is: {\em bosonization}, a duality between two quantum field theories in two spacetime dimensions: one with bosons, and one with fermions. So here: a bosonic model triple is isomorphic (by a unitary equivalence) to a fermionic model triple, their common core being a certain infinite-dimensional algebra (2017: Sections 4, 5). Given how different bosons and fermions are, this is a remarkable, indeed very surprising, isomorphism. But as it happens: it is {\em not} a case of (Contr) or (Diff).

\subsection{Details of the Schema: duality as isomorphism of model triples}\label{details}
It is clearest to have a notation in which the model triple, comprising the state-space $\cal S$, the set of quantities $\cal Q$, and the dynamics $\cal D$:  (i) is separated from the model's own specific structure, and (ii) expresses only the model's realizing (typically: representing in the mathematical sense) the bare theory $T$. Thus we write a model $M$ of $T$ as
\beq
 M=\bra{\cal S}_M,{\cal Q}_M,{\cal D}_M,\bar M\ket  =:\bra m,\bar M\ket~,
\eeq
 where: $m :=  \bra {\cal S}_M, {\cal Q}_M, {\cal D}_M\ket$ is the model triple; and $\bar M$ is the {\em specific structure}, the  `details', that a model adds to its bare theory. 
 
 But note that $m$ is usually built by using the specific structure $\bar M$. So it should not be thought of as given independently of $\bar M$; and one should not think of $M$ as just the ordered pair of independently given $m$ and $\bar M$. Rather, $m$ encodes $M$'s representing the bare theory $T$. So one might well write $T_M$ instead of $m$, since having a subscript $M$ on the right hand side of the equation $M = \bra T_M, \bar M \ket$ signals that $M$ is not such an ordered pair. In other words: the notation $T_M$ emphasises that the model triple: (i) is a representation of $T$, (ii) is built from $M$'s specific structure viz. $\bar M$, and (iii) is itself a theory (hence the letter `T').
 
To repeat Section \ref{idea}'s warning against misleading connotations of the word `model': the specific structure or details, $\bar M$, are {\em not} a matter of specifying: (i) a solution or history of the system; or (ii) approximation(s); or (iii) interpretation(s). Rather, the extra details are extra mathematical structure: just like a representation of a group or an algebra has extra details or structure, beyond that of the group or  algebra of which it is a representation.\\

It is clear that two model triples that model a bare theory are in general {\em not} isomorphic to each other, nor to the bare theory: in the relevant sense of `isomorphic', viz. given by the bare theory. (The situation is just like in representation theory.) So the assertion of duality is substantive: it asserts that two model triples are in fact isomorphic. \\
 
Before formally defining duality as isomorphism, we sketch how states, quantities and dynamics will be treated. Suppose we are given a set of states ${\cal S}$, a set of quantities ${\cal Q}$ and a dynamics ${\cal D}$: $\bra {\cal {S}}, {\cal {Q}}, {\cal {D}} \ket$. We will write $\langle Q , s \rangle$ for the value of quantity $Q$ in state $s$. It is these values that are to be suitably preserved by the duality, i.e. by the isomorphism of model triples: cf. eq. \ref{obv1} below.
 
For classical physics, one naturally takes quantities as real-valued functions on states, so that $\langle Q , s \rangle := Q(s) \in \mathR$ is the system's possessed or intrinsic value, in state $s$, of the quantity $Q$. For quantum physics, one naturally takes quantities as linear operators on a Hilbert space of states, so that  $\langle Q , s \rangle := \langle s |{\hat Q} | s \rangle \in \mathR$ is the system's Born-rule expectation value of the quantity. (But in fact, for quantum dualities, the duality often preserves off-diagonal matrix elements $\langle s_1 |{\hat Q} | s_2 \rangle \in \mathC$: cf. below.)  
 
 As to dynamics, here assumed deterministic:--- This can be presented in two ways, for which we adopt the quantum terminology, viz.~the `Schr\"{o}dinger' and `Heisenberg' pictures; (though the ideas occur equally in classical physics). We will adopt for simplicity the Schr\"{o}dinger picture. So we say: $D_S$ is an action of the real line $\mathR$ representing time on ${\cal S}$. 
 
 There is an equivalent Heisenberg picture of dynamics with $D_H$,  an action of  $\mathR$ representing time on ${\cal Q}$. The pictures are related by, in an obvious notation:
\be
D_S:   \mathR \times {{\cal S}} \ni (t,s) \mapsto D_S(t,s) =: s(t) \in {{\cal S}} \;  {\mbox{iff}} \; 
D_H:   \mathR \times {{\cal Q}} \ni (t,Q) \mapsto D_H(t,Q) =: Q(t) \in {{\cal Q}} 
\ee 
where for all $s \in {{\cal S}}$ considered as the initial state, and all quantities $Q \in {{\cal Q}}$, the values of physical quantities at the later time $t$ agree in the two pictures:
\be
\langle Q, s(t) \rangle = \langle Q(t), s \rangle \; .
\ee \\

We can now present our Schema for duality as an isomorphism between model triples. Let $M_1, M_2$ be two models, with model triples $m_1 =  \bra {\cal S}_{M_1}, {\cal Q}_{M_1}, {\cal D}_{M_1}\ket$ and $m_2 =  \bra {\cal S}_{M_2}, {\cal Q}_{M_2}, {\cal D}_{M_2}\ket$. We can suppose that  $M_1, M_2$ are both models of a bare theory $T$. 

To say that the model triples $m_1, m_2$ are isomorphic is to say, in short, that: there are isomorphisms between their respective state-spaces and sets of quantities, that \\
\indent \indent (i) make values match, and \\
\indent \indent (ii) are equivariant for the two triples' dynamics (in the Schr\"{o}dinger and Heisenberg pictures, respectively).\\
Thus these maps are our rendering of physics jargon's dictionary.\\

Thus we say:--- A {\bf duality} between $m_1 =  \bra {\cal S}_{M_1}, {\cal Q}_{M_1}, {\cal D}_{M_1}\ket$ and $m_2 =  \bra {\cal S}_{M_2}, {\cal Q}_{M_2}, {\cal D}_{M_2}\ket$ requires: \\
\indent \indent (1): an isomorphism between the state-spaces (almost always: Hilbert spaces, or for classical theories, manifolds): 
\be\label{obv-1}
d_s: {{\cal S}_{M_1}} \rightarrow {{\cal S}_{M_2}} \;\;  {\mbox{using $d$ for  `duality'}} \; ; \; {\mbox{and}}
\ee 
\indent \indent (2): an isomorphism between the sets  (almost always: algebras) of quantities\\
\be\label{obv0}
d_q: {{\cal Q}_{M_1}} \rightarrow {{\cal Q}_{M_2}} \;\;  {\mbox{using $d$ for `duality'}} \; ;
\ee 
such that: (i) the values of quantities match: 
\be\label{obv1}
\langle Q_1, s_1 \rangle_1 = \langle d_q(Q_1), d_s(s_1) \rangle_2 \; , \;\; \forall Q_1 \in {{\cal Q}_{M_1}}, s_1 \in {{\cal S}_{M_1}}. 
\ee
and: (ii) $d_s$ is equivariant for the two triples' dynamics, $D_{S:1}, D_{S:2}$, in the Schr\"{o}dinger  picture; and 
$d_q$ is equivariant for the two triples' dynamics, $D_{H:1}, D_{H:2}$, in the  Heisenberg picture: see Figure.\\

\begin{figure}[ht]
\begin{center}
\bea
\begin{array}{ccc}{\cal S}_{M_1}&\xrightarrow{\makebox[.6cm]{$\sm{$d_s$}$}}&{\cal S}_{M_2}\\
~~\Big\downarrow {\sm{$D_{S:1}$}}&&~~\Big\downarrow {\sm{$D_{S:2}$}}\\
{\cal S}_{M_1}&\xrightarrow{\makebox[.6cm]{\sm{$d_s$}}}&{\cal S}_{M_2}
\end{array}~~~~~~~~~~~~
\begin{array}{ccc}{\cal Q}_{M_1}&\xrightarrow{\makebox[.6cm]{$\sm{$d_q$}$}}&{\cal Q}_{M_2}\\
~~\Big\downarrow {\sm{$D_{H:1}$}}&&~~\Big\downarrow {\sm{$D_{H:2}$}}\\
{\cal Q}_{M_1}&\xrightarrow{\makebox[.6cm]{\sm{$d_q$}}}&{\cal Q}_{M_2}
\end{array}\nonumber
\eea
\caption{Equivariance of duality and dynamics, for states and quantities.}
\label{obv2}
\end{center}
\end{figure}

Eq. 6  appears to favour $m_1$ over $m_2$; but in fact does not, thanks to the maps $d$ being bijections.\\

 De Haro and I have related this Schema, in some detail, to various topics such as (a) the idea of symmetry, (b) the construction of theories  by abstraction from models,  and (c) the interpretation of theories; (cf.  2017,  Sections 3.1, 2.4, 2.3, 3.2.3.) For  this paper, I only need to review (a) the idea of  symmetry (Section \ref{symmthy}) and (c) the interpretation of theories (Section \ref{subject}). The latter will make precise the idea of a subject-matter; and this will lead in to presenting examples of the Remark (Section \ref{weak;exples}). 
 
 \subsection{Symmetries of a theory: self-dualities}\label{symmthy}
The main connection between this Schema and the idea of symmetry is straightforward. For if  we set aside the bare theory, and the specific structure of the two models, we have, up to isomorphism, just one model triple. Therefore, the conditions (i) and (ii) in   Section \ref{details}'s definition of duality become the usual definition of a symmetry of a theory: i.e. a dynamical symmetry, that transforms a dynamically possible history into another `replica' history.\footnote{Cf. 2017, Sections 3.1, 3.2.4. The latter, and that paper's item (4) in its Section 2.1, compare dualities with the idea of `gauge'. Note that  Rickles (2017) argues that dualities are strongly analogous to `gauge'.}  

But there are two further points to make. (1): The first is to note that we often regard a state and its transform under a symmetry as physically distinct. This is a very simple point, and I will illustrate it with a very simple example: rotation in space. But it clearly illustrates the broad theme that will underpin the Remark: viz., items can have the same description but be physically distinct. (2): The second  point concerns the jargon of `self-duality', and illustrates the same theme.\\

(1): Consider the rotational symmetry of Newtonian point-particle mechanics with a rotationally invariant Hamiltonian (say, $N$ particles mutually gravitating). Thus if we rotate a configuration about an axis by some angle, we get another configuration: it is a `replica', apart from orientation in space. Similarly, if we rotate a state (i.e. a configuration together with velocity/momentum information): we get another state; and  thanks to the determinism towards both past and future, we get a rotated total history (where `history' means trajectory through the state-space)---a `replica' of the unique history through the original state. (Of course, an arbitrary translation through space give a similar example, assuming a translationally invariant Hamiltonian.)

This example will be covered in more general terms in Section \ref{weak;exples} (Example (1) and, in fancier language, Example (5)).  But we can already see the point. Namely: it is {\em not} mandatory to say that rotation is `gauge', that there should be no physical difference between the original state/history and the rotated one. Agreed: you {\em might} wish to say this, especially if the $N$ particles are the entire material contents of the universe, so that the rotation of all of them is not observable by any means  (since there are no other bodies in the universe). Here, we are of course back to the debate between absolute and relational conceptions of space: paradigmatically between Newton and Clarke advocating the absolute conception, and Leibniz advocating the relational conception.\footnote{Of course, when there are other bodies, both sides agree that the rotation is not `gauge'.} 

I do not need to adopt any view in this debate. I only need to point out that neither side has been agreed to be the victor. Thus it is not mandatory to `join Leibniz'.  In other words: Newton's and Clarke's views are tenable. And according to them: here is a case of physically distinct states with, in a perfectly good sense, `the same description'. \\

(2): A symmetry of a theory is sometimes called a   {\em self-duality}: though this label is not used for spacetime symmetries like rotational or translational symmetry of Newtonian  gravitation. Two examples where the label {\em is} used are: (i) the classical harmonic oscillator in one spatial dimension, and (ii) classical vacuum electromagnetism. It will be enough to present these informally, without specifying precisely the sets ${\cal S}, {\cal Q}$ of states and quantities. Again, my point will be that here we have unproblematically distinct states---in my jargon: `the duals disagree'.  (As I mentioned in footnote \ref{WAS2qualif}, other authors agree on these examples: Matsubara (2013: 478), Huggett (2017: 86) and Dieks et al. (2015: 209).)

\indent (i): The harmonic oscillator, with $H = \frac{p^2}{2m} + \frac{1}{2}m \omega^2 x^2$. Define $d \equiv d_s$ on the $(x,p)$ plane $\mathR^2$ by
\be
d: x \mapsto \frac{p}{m \omega}  \;\; ; \;\; p \mapsto - m \omega x \; . 
\ee
Then $d(H) = H$, i.e. $d$ preserves the ellipses of equal energy in the $(x,p)$ plane: $d$ combines a clockwise rotation by $\frac{\pi}{2}$ with a re-scaling (viz. dilation by $m \omega$ and contraction by $\frac{1}{m \omega}$).\\
\indent But NB: there is no temptation to identify a state $\langle x, p \rangle$ and its image $d( \langle x, p \rangle )$. If  describing {\em this} harmonic oscillator as in $\langle x, p \rangle$ at time $t_0$ is accurate, then describing it as in $d( \langle x, p \rangle )$ is plain wrong!\\
\indent   So again, this is a case of: `Newton and Clarke's views are tenable'. Indeed, I would say that they are tenable, even if the harmonic oscillator was the only system in the universe. That is: it is tenable to say, for a `lonely' harmonic oscillator at a certain time, that if describing it as in $\langle x, p \rangle$ at time $t_0$ is accurate, then describing it as in $d( \langle x, p \rangle )$ would be plain wrong.

(ii): In classical vacuum electromagnetism, the map 
\be
d: E \mapsto - B \;\; ; \;\; B \mapsto  E \; . 
\ee
maps solutions of Maxwell's equations to solutions, i.e. is a symmetry of the theory. (It is also `squares to minus 1', i.e. $d^2 = -1$.) \\
\indent But again: `Newton and Clarke's views are tenable'. It is not mandatory to say there is no physical difference between an  argument-solution, i.e. an assignment of $E$ and $B$ throughout Minkowski spacetime $\cal M$, and the image-solution defined by applying $d$. In the argument-solution, here at  $p \in {\cal M}$, there is, say, ${\bf E}(p) = (5,4,3) \in \mathR^3$ in some cartesian coordinates; in the image-solution, instead, we have  ${\bf B}(p) = (-5,-4,-3) \in \mathR^3$ in those cartesian coordinates. That looks to be a physical difference.\footnote{Again: I agree that these `absolutist' views, though tenable, can be denied. They stand accused of drawing a distinction where there is no difference: more precisely, of claiming, controversially, that there are, or can be, in-principle-unobservable differences. In philosophy, the claims have labels. Claiming such differences between objects, say the spatial points in (1)'s rotation example, is called {\em haecceitism}. Claiming such differences between properties, say the electric and magnetic fields in example (2: ii), is called  {\em quidditism}: a claim that is less often denied than is haecceitism---but with arguments to confront (Black 2000). As I said at the end of Section \ref{others}: I take up the discussion in Section \ref{Ex1} and in (5) at the end of Section \ref{Ex6}.}\\


\section{Interpreting physical theories}\label{subject}
`Interpreting physical theories' is a large subject, at the intersection of logic, especially semantics, and philosophy of science. In this Section, I undertake three tasks. I report my endorsement of {\em intensional semantics} (Section \ref{intsemics}). Then I report how that framework makes precise the notion of {\em subject-matter}---and thereby, Section \ref{intro}'s two `ways to disagree', (Contr) and (Diff) (Section \ref{s-matters}). The ``shape'' of my Remark will then be clear (Section \ref{credo}).

\subsection{Intensional semantics}\label{intsemics}
This is the framework for analysing the meanings of words and sentences, developed by such authors as Carnap, Montague and Lewis, that combines Frege's distinction between sense and reference with the tradition, from Leibniz to modern modal logic, of a set of possible worlds, i.e. all the ways the universe could be, only one of which is actual. 

At its simplest: a singular term such as `the Earth's tallest building' has an intension (a Fregean sense) that is a partial function on the set $W$ of worlds mapping a world $w \in W$ to the term's extension (reference) in $w$---in the example, to the tallest building on Earth, in $w$. And if $w$ has no Earth, or Earth in $w$ has no buildings, then the function is undefined. Similarly, a predicate such as `is made of steel' has an intension that is a function sending each world $w$ to the predicate's extension, i.e. set of instances, in $w$. The semantics of sentences, both simple and compound, is then built up compositionally.  The intension of a sentence $S$ is defined as the function sending a world $w$ to 1 or 0, according as $S$ is true or false at $w$, as determined by the extensions at $w$ of $S$'s parts.  So the intension of $S$ is the characteristic function of the set of worlds where $S$ is true; and the extension of  $S$ at  $w$ is 1 or 0. Or equivalently: the intension of a sentence is the set of worlds where it is true. Thus the intension of `and' is the set-theoretic operation of intersection among sets of worlds. And so on.

Although this framework was originally developed for formal languages in logic and for (fragments of!) natural languages, it is readily adapted to physical theories with their talk of systems, states and quantities, as in Section \ref{details}. Roughly speaking: systems are objects, i.e. references of singular terms; quantities are relations of objects to numbers (relative to a system of units); a state is a value-assignment to all the quantities that apply to a given object; and a possible world is the total histories of the states of all the objects in the world.

De Haro and I endorse this framework, especially in the version by Lewis (1970). And our endorsement is not just in general: but as applied to Section \ref{details}'s conception of theories, models and dualities. Thus the assignment of intensions and extensions to states and quantities in model triples can be made precise in terms of {\em interpretation maps}. These  are partial functions on the sets of states and quantities, $\cal S$ and $\cal Q$, into sets of intensions (Fregean senses) or extensions (worldly referents). Details are given in (2017, Section 2.3; and as applied to dual model triples, Section 3.2.3).  But this paper does not need further details about this framework, except in two regards. 

(1):  The framework has the great merit of respecting the  meanings of words! It takes  there to be a fact about what is the correct intension of  `the Earth's tallest building': relative of course to various factors, such as choice of language, resolutions of vagueness, context of utterance etc. That may seem to you a feature so obviously mandatory for an endeavour calling itself `semantics', that you may ask why I have stressed it. But as we will see in Section \ref{logic}, there are `semantics' that investigate the mathematical consequences of assigning arbitrary meanings (specifically, extensions) to words---viz. the semantics in  books of logic, and especially model theory!

Two qualifications about this trenchant remark `against logic'. (i): It will become clear in Section \ref{logic} that this comment is meant as a cheeky appetizer for my argument, rather than as a criticism of those books. Of course, logic and model theory teach us a great deal about semantics, broadly understood, by studying the consequences of assigning arbitrary meanings. (ii): I agree that it is a good question, perhaps the central question of philosophy of language, exactly what facts make true the correct assignment of  intensions. About this,  I admire Lewis' answer (1974, 1975, 1983: 370-377). I also admire Callender and Cohen's claim that we should analyse representation in science in terms drawn from philosophy of language and mind (2006, especially Section 3). I also maintain that these authors' claims are compatible with Section \ref{prosp}'s two morals.

(2): Later on, it will be useful to deploy the framework's  precise notion of a subject-matter, as a partition of the set $W$ of worlds: to which I now turn. 

\subsection{Subject-matters: (Contr) and (Diff)}\label{s-matters}
As so often, one cannot say it better than Lewis. He writes:
\begin{quote}
We can think of a subject matter, sometimes, as a part of the world:
the 17th Century is a subject matter, and also a part of this world.
Or better, we can think of a subject matter as a part of the world {\em in
intension}: a function which picks out, for any given world, the
appropriate part---as it might be, that world's 17th Century. (If for
some reason the world had no 17th Century, the function would be
undefined.) We can say that two worlds are exactly alike with
respect to a given subject matter. For instance two worlds are alike
with respect to the 17th Century iff their 17th Centuries are exact
intrinsic duplicates (or if neither one has a 17th Century).

This being exactly alike is an equivalence relation. So instead of thinking of a subject matter as a part of the world in intension, we can think of it instead as the equivalence relation. This seems a little artificial. But in return it is more general, because some subject matters---for instance, demography---do not seem to correspond to parts of the world. [...]

The equivalence relation on worlds partitions the worlds into
equivalence classes. The equivalence classes are propositions, ways
things might possibly be. An equivalence class is a maximally
specific way things might be with respect to the subject matter. So a
third way to think of a subject matter, again general, is as the
partition of equivalence classes. (1988: 111-112). 
\end{quote}
Thus the idea is that a subject-matter is a taxonomy or classification-scheme, so that worlds that match exactly as regards the taxonomy are to be in the same cell of the partition of $W$ (the same equivalence class). Lewis goes on to discuss various ensuing ideas and applications (1988: 112-117; cf. also 1988a: 136-141). I need only note three of the ideas, and can set aside Lewis' applications, interesting though they are. These three ideas will enable me---after a brief skirmish with the quantum measurement problem---to state (Contr) and (Diff) more precisely. \\

First: one subject-matter can {\em contain}, i.e. be more detailed than, another: like the 17th Century contains the 1680s, or the subject-matter, how many stars there are, contains the subject-matter, whether there are finitely or infinitely many stars. In terms of partitions, this is a matter of the first partition being a refinement, or fine-graining, of the second. That is: each cell of the second partition is a union of cells of the first. Thus the bigger subject-matter, with more cells in its partition, is a refinement of the smaller. Thus the set of subject-matters is partially ordered by `is a refinement of'. And in this partially ordered set, one can consider least upper bounds and greatest lower bounds: i.e. the `finest common coarse-graining' and the `coarsest common fine-graining' of a set of subject-matters. Of course, subject-matters might form a very sparse subset of the set of all partitions of $W$. So there need be no commitment to such bounds: in particular, to a greatest lower bound of all subject-matters. And even if that greatest lower bound exists, it need not be so dizzyingly fine as the set of all singletons of worlds in $W$. 

Second: a proposition is {\em about}, i.e. entirely about, a subject-matter if its truth-value is determined by the facts about the subject-matter: that is, if the set of worlds at which it is true is a union of cells of the subject-matter. So if a proposition is about a subject-matter, it is also about any bigger (more detailed, finer-grained) subject-matter. So we cannot speak unequivocally of {\em the} subject-matter of a proposition.

Third: In general, of course, partitions {\em cut across} one another, in that neither is a refinement of the other. That is: at least one cell of one partition overlaps, i.e. intersects, two cells of the other. The extreme case, which Lewis calls the subject-matters being {\em orthogonal}, is: each cell of each partition intersects each cell of the other. An example might be: how many stars there are, as a subject-matter, is orthogonal to Queen Victoria's preferences about clothes. That is: any number of stars is compatible with any sartorial tastes she could have had. \\

 It is clear that even if you are suspicious of possible worlds, and-or of intentional semantics,  these ideas will be useful whenever you are concerned with various partitions of some given set. For a philosopher of physics, the prototype is of course partitions of the phase space $\Gamma$, in classical mechanics. Each quantity $Q$, taken as a real function on $\Gamma$, defines a partition of $\Gamma$ whose cells are its level surfaces (inverse images of $Q$'s values). So $Q$ defines a subject-matter.  Taking a function, $f(Q)$, of $Q$ is, in general, a coarse-graining. For in general, some two arguments of a function $f$ yield the same value: so that at least one cell of $f(Q)$'s partition is a union of cells from $Q$'s partition. This prototype also gives examples of orthogonality in Lewis' sense. For many classical mechanical systems, any value of position for some given component of the system is compatible with any value of its momentum, so that the component's position and momentum are orthogonal as subject-matters. Similarly, for the same quantity---position, say---on two different components: each component can be anywhere, wherever the other is.

Similarly neat examples of subject-matters, and of these related ideas, will be evident in  Section \ref{weak;exples}'s elementary examples of dualities. Agreed, there will be the usual elephant in the room: viz. the {\em quantum measurement problem}. More precisely: quantum theory's superposition principle, or non-commutativity of quantities. I shall set this aside; but we should register the issues here.

Clearly, the ideas of a subject-matter as a partition of the possible worlds, and of a quantity defining a partition of a classical  phase space,  fit well together. They are `Boolean cousins'. But adapting the latter to quantum physics yields the quantity's spectral family, i.e. its complete mutually orthogonal family of eigenspaces (equivalently: the corresponding projectors). This raises the question how all the countless superpositions `between' different eigenspaces are to be included in any putative set of `worlds'. More pointedly: in some quantum cases we will later classify as `different subject-matters', e.g. position and momentum (cf. Section \ref{Ex2}), the spectral families `cut across' one another in a more radical, even weird, way than Lewis' orthogonality, just mentioned. Namely: for any two real intervals $[a_1, b_1], [a_2, b_2] \subset \mathR$, the position projector for $[a_1, b_1]$, $E^Q_1$ say, and the momentum projector for $[a_2, b_2]$, $E^P_2$ say, have as their meet, $E^Q_1 \wedge E^P_2$, i.e. the intersection of their ranges, the {\em zero subspace}. That is: no state can be of compact support in both position and momentum. So here, the `cutting across' is not just a matter of a cell of one partition overlapping two from the other, i.e. an eigenspace of one family having non-zero projections on two eigenspaces of the other family.  

But these issues  will make no difference to the arguments, and conclusions, of this paper; and so I will not pause over them. All I need is to use the above summary of  intensional semantics and subject-matters, so as to give a more precise statement of Section \ref{intro}'s two ways that two theories can disagree,  (Contr) and (Diff). Though these formulations ignore the issues just raised, they will be precise enough for my purposes---and they will dominate the rest of the paper.   \\

In Section \ref{intro}, I said that for (Contr): two dual theories make contrary assertions about a common subject-matter, so that though they may agree on some claims, they contradict one another over other claims. Given Section \ref{details}'s Schema, and the above account of possible worlds, propositions and subject-matters: we think of conjoining the assertions made by an interpreted model triple into a single proposition, which is made true by a set of worlds. Since the interpreted model triple is about an intended subject-matter (such as---to look ahead to Section \ref{Ex1}---point particles, interacting by Newtonian gravity), this set of worlds is a union of cells of the partition that {\em is} the subject-matter.  Thus   we have:\\
\indent \indent (Contr):  Each of two dual model triples is interpreted as wholly true (its conjunctive proposition is wholly true) at a union of cells of a common	subject-matter. But these two unions are disjoint: for the propositions contradict each other. \\

Similarly: in Section \ref{intro}, I said that for (Diff): two dual theories describe different subject-matters, though the descriptions are `isomorphic'  or  `matching'---hence the duality. Applying again the notions introduced above, this becomes: the two interpreted model triples are about two different subject-matters (such as---to look ahead to Sections \ref{Ex2} and \ref{Ex3}---position as against momentum, or low temperature as against high temperature). And these subject-matters, i.e. partitions, cut across one another: maybe even in the extreme sense Lewis dubbed `orthogonality'. Thus we have:\\
\indent \indent   (Diff): Two dual model triples are interpreted as wholly true (the conjunctive proposition of each is wholly true)  at distinct sets of worlds. Each set is a union of cells of the triple's subject-matter, i.e. partition. But the partitions are different, and so are the sets. The sets need not be disjoint: both the model triples could be, both of them, wholly true. But the sets are distinct.\\

So much by way of making subject-matters, and thereby (Contr) and (Diff), more precise. As I mentioned: for most of the paper, this precision is not needed. In particular, one can work with Section \ref{intro}'s formulations of (Contr) and (Diff). But the precision will be helpful when we come to string theory (Section \ref{string}). 

\subsection{Credo}\label{credo}
The ``shape'' of my Remark is now clear. By invoking the intended meanings or interpretations of physical theories, I will exhibit cases of (Contr) and (Diff). The isomorphisms given by the duality (eq. \ref{obv-1} to \ref{obv1}) will not be ``allowed'' to rule out such cases by their ``carrying'' meanings across to one dual from the other, i.e. their fixing the first dual's interpretation by leaning on the isomorphism so as to use an existing interpretation of the second dual. For carrying meanings across in that way would violate the first dual's intended meaning.

Of course, I do not mean to suggest that two duals' intended meanings are usually, let alone always, different: not all pairs of duals disagree, in my umbrella sense. Indeed: since Section \ref{details}'s definition of duality is formal, regardless of meanings, some pairs agree ``utterly''---agree in a stronger sense than do the duals in position-momentum duality, or in bosonization. Thus consider a formulation of relativity theory with a $(+,-,-,-)$ signature for the metric, and another formulation  with a $(-,+,+,+)$ signature. (Thanks to Thomas Barrett for pressing this example.) {\em Prima facie}, the first formulation says that a particle's worldline has positive length, the second that it has negative length. But of course, everyone concurs that these formulations in no way disagree. The $+/-$ in the metric's signature is a paradigm case of a `trivial convention'; and so the two formulations, construed with their intended meanings, are indeed equivalent. 

Should we then try to strengthen the definition of duality, so that such cases of ``trivial equivalence'' are ruled out as not being dualities? I resist this project, because of the first moral in Section \ref{prosp}. The project would require distinguishing `fact' from `convention' once and for all, across a whole range of theories (or if you prefer: formulations of theories).  And since at least sixty years ago, that has seemed like a mug's game. So: better, say I, to keep to a formal definition of duality, independent of interpretations; and to admit that when classifying an example of it as the duals agreeing (`equivalent'), or as disagreeing ((Contr) or (Diff)), we must exercise judgment about what best respects each dual's intended meaning. 

Similar remarks apply in connection with the fact that our Schema is not just formal, but logically weak. So it fits countless cases which would not normally be called `dualities'. And countless such cases would not even be articulated at all---we need no special word for isomorphic representations of any algebraic structure!  

But  De Haro and I think (2017, Sections 2.1, 3.2.2) that it is best to keep to a formal and logically weak definition of `duality', while admitting that of course, the {\em scientific interest} of a duality depends on interpretative issues: in particular, the isomorphism having such features as the following:\\
\indent \indent (i)  being non-obvious, and-or \\
\indent \indent (ii)  being of a logically strong structure (e.g. `10 dimensional Lie group', not just `group'), and-or \\
\indent \indent (iii) having other scientific advantages: one important case  being that the duality renders the weak coupling, and therefore tractable, regime of one theory dual to the strong coupling, and therefore intractable, regime of the other---so that one can solve `easy' problems in the tractable regime of the first theory so as to solve `hard' i.e. intractable problems in the second theory.\footnote{These three desiderata are of course widely endorsed, though often with different labels: like `Striking' and `Useful' in De Haro, Teh, and Butterfield (2017: Section 2)). Thus the famous duality, bosonization, on which our (2017) concentrated, has all of them.  Here, we return to dualities' {\em heuristic} importance for guessing new theories, which I mentioned at the end of Section \ref{others}: and will take up again in Sections \ref{Ex1} and \ref{Ex6}.}  

Agreed: one might want to strengthen the definition of `duality' by requiring such features. As we shall see, feature (i) is often a matter of the notions that correspond by the duality---say, the quantity $Q$ on the `left', $Q \in {\cal Q}_{M_1}$, and its image $d_q(Q)$ on the `right'---being {\em disparate}. (Recall Section \ref{idea}'s comment that a duality `dictionary' need not be faithful to meanings.) Thus: in one of Section \ref{weak;exples}'s examples, $Q$ is position, while $d_q(Q)$ is momentum; and in another example, $Q$ is free energy at a low temperature, while $d_q(Q)$ is free energy at a  corresponding high temperature.

 But it is impossible to be precise about amounts of disparateness, or non-obviousness: or of logical strength. So again: I resist this project of strengthening the definition, because of the morals in Section \ref{prosp}.  In any case, this paper's project does not need any such strengthening: though, as we shall see, the examples that exemplify the Remark have (or, as a matter of history: once had) most of the features (i) to (iii).

So much by way of a {\em credo} about what is worth making make precise (even formal), and what is best left to judgment in each case.  I now turn to elementary examples of the Schema ...

\section{Examples in classical and quantum physics}\label{weak;exples}
In this Section, I present five  {\em Examples}, (1) to (5), of the Schema. Of these, just two, viz. (2) and (3), are usually called `dualities'. And I will concentrate on the first three, which exemplify the Remark. The Remark will be evident: two dual theories can disagree,  in one of the two ways, (Contr) and (Diff). They can contradict one another about the same subject-matter, or they can be about different subject-matters (indeed: orthogonal subject-matters, in Lewis' sense).

I will only sketch these Examples.  There will be two main limitations of scope. First: it will be clear that there are choices to be made about what exactly is the bare theory, and thus the notion of representation in play in defining the models/duals, i.e. in defining the model triples and their specific structure. But we will not need to make very precise choices, in order to illustrate the Schema adequately: and in particular, to exemplify the Remark. Second: I will downplay the role of dynamics, i.e. eq. \ref{obv2}: partly to save space, and partly because in some Examples, dynamics  `drops out' from the discussion. (For more detail about these Examples, cf. Butterfield and De Haro (2018).) 

But although the Examples are only sketched, one can check that for each of them, most or even all of Section \ref{credo}'s features (i) to (iii) about {\em scientific interest} hold---or, as a matter of history, once held. That is: the equivalence/duality: \\
\indent \indent (i) is non-obvious (or rather: it was non-obvious at a certain historical epoch!); \\
\indent \indent (ii) is of logically strong structures (or rather: the structures seemed logically strong  at a certain historical epoch---before the further development of physics made, e.g. the Fourier transform between position and momentum, look elementary);\\
\indent \indent (iii) has scientific advantages, e.g. by one of the duals being useful when the other is not---though, as it happens, only in Example (3) is the scientific advantage a matter of pairing weak and strong couplings (more precisely: temperatures).

\subsection{(1):  Newtonian mechanics with different standards of rest}\label{Ex1}
Consider two formulations of Newtonian point-particle mechanics (say $N$ particles with  gravitation), that differ in what they identify as absolute space (what inertial timelike congruence is `truly at rest'). 

Here the natural choice for the bare theory would be a neo-Newtonian (also called: `Galilean')  formulation of point-particle mechanics, which postulates a flat affine connection on spacetime, but no absolute rest. Thus the spacetime might be taken to be $\mathR^4$. Then in the bare theory, the  obvious congruence of inertial timelike lines, i.e. the lines $\langle t, x, y, z \rangle$ (with  $t$ varying, and $x,y,z$ fixed, for each line) is on a par with---no more `at rest' than---any congruence boosted with respect to it, such as $\langle t, x - vt, y, z \rangle$ (with  $t$ varying, and $x,y,z$  fixed, for each line; and $v$ fixed for the whole congruence; representing a boost in the $x$ direction).  

Then the two duals (two models, in our sense) would be formulations of point-particle mechanics set in Newtonian spacetime. So the specific structure by which the two duals differ lies in their specifications of absolute rest. The `left' dual might specify as absolute rest, the first-mentioned congruence, with lines $\langle t, x, y, z \rangle$; while on the other hand,  the `right' dual specifies  $\langle t, x - vt, y, z \rangle$.

Let us fill this out a little in terms of the state-space, the map $d_s$ etc. The configuration space of the $N$ particles can be taken to be $\mathR^{3N}$. (I say `can be taken' because (i) one might instead take the affine space, and (ii) one might remove the coincidence points where two or more particle locations coincide.) Then a state can be an assignment of values of (absolute!) position and velocity to each particle: $(x_1, ... z_N; {\dot x_1},....{\dot z_N})$. We may then define the duality map $d_s$ on states so that it takes a state $(x_1, ... z_N; {\dot x_1},....{\dot z_N})$ of the first (`left') model to the state in the second (`right') model  with the same numerical values, with respect to {\em its} (the right model's) specification of absolute rest, as the argument-state had for the the left model's specification. This definition of $d_s$, together with a definition of $d_q$ that sends `absolute rest' to `absolute rest' etc.   will preserve the values of absolute quantities, i.e. would satisfy eq. \ref{obv1}.

Agreed: there is another, equally natural, way to define $d_s$ and $d_q$. Namely: $d_s$ is to take a state $(x_1, ... z_N; {\dot x_1},....{\dot z_N})$ of the left model to the  same  state `as regards $\mathR^4$'  (of course, as described by the right model). Here, `same as regards $\mathR^4$' means that if Mr Right's absolute rest is boosted with respect to Ms Left's by, say, velocity $v$ in the $x$ direction, then particles that are at rest according to Ms Left, ${\dot x_1} = {\dot y_1} = ... = {\dot z_N} \equiv 0$ will be moving at  velocity $- v$ in the $x$ direction according to Mr Right, i.e. with primed coordinates ${\dot x_1}' = {\dot x_2}' = ... = {\dot x_N}' \equiv - v$: and so $d_s$ is to fix their worldlines taken as (unprimed) coordinate curves in $\mathR^4$. (Again,  the paper's {\em motto} applies!) For brevity, I will lean on the structure of $\mathR^4$ with its distinguished coordinate lines, and say that this $d_s$ fixes the {\em physical} state.   But however one labels (or thinks of) this $d_s$, it still allows the duality to preserve values, i.e. satisfy eq. \ref{obv1}, by our suitably defining $d_q$ to `compensate for the boost'. \\

For us, the important point is that, with either the first or the second way of defining the duality maps $d_s$ and $d_q$, this Example is a case of (Contr). For the duals disagree about what is absolute rest. 

On the first way of defining the duality maps, the values match, i.e. eq. \ref{obv1} is satisfied, even though $d_q$ pairs Ms Left's and Mr Right's intended meanings for `absolute rest'---and so respects their mutual disagreement, so that the reference, or extension, of `absolute rest' is different on the two sides. Namely: The values match, eq. \ref{obv1} is satisfied, because $d_s$ is {\em not} the identity map on what I called the `physical states': $d_s$ maps a state of Ms Left to another state that gets the same numerical values for Mr Right's corresponding (by $d_q$) quantities. 

And on the other hand: on the second way of defining the duality maps, the values match, i.e. eq. \ref{obv1} is satisfied, even though now $d_s$  {\em is} the identity map on physical states. Namely: The values match, eq. \ref{obv1} is satisfied, because now $d_q$ is defined so as to {\em not} pair Ms Left's and Mr Right's intended meanings for `absolute rest'---but rather, to `compensate for the boost'.\\

To sum up: The `left' isomorphic set of states is all the possible instantaneous states of $N$ particles that inhabit a Newtonian spacetime, equipped with `this' absolute rest ; and the `right' isomorphic set of states is all the possible instantaneous states of $N$ particles that inhabit a Newtonian spacetime, equipped with `that' absolute rest, defined as boosted with respect to the left, i.e. `this'  one. (I spoke in terms of instantaneous states; but given the two-way determinism, I could have said `all the possible histories'.) And this pair of duals, on the left and right, illustrates (Contr).\footnote{Agreed: my discussion of spacetime, with its dubbing states `physical' in terms of the structure of $\mathR^4$, is a `cartoon'. (Thanks to T. Barrett, G. Leegwater and O. Pooley for pointing this out.) For a proper discussion, cf. e.g. Malament (2012, Chapter 4), Pooley (2019, Chapter 4.3, 4.4). But despite the cartoon, the point remains that the empirical undetectability of Newton's absolute space can be rendered in spacetime terms as two formulations disagreeing about {\em which} timelike congruence is at rest, and thus exemplifying (Contr).} \\

{\em Beware}: it is tempting to say that there is no real disagreement,  that the contrary specifications of absolute rest are `gauge', or `a distinction without a difference', or `a sign that we should move to a neo-Newtonian formulation of point-particle mechanics', a formulation in which the `surplus structure  of absolute rest (i.e. specific structure in our Schema's jargon) is eliminated'. 

I agree that it is {\em tempting} to say these things. But the point is: these temptations are the benefit of hindsight, i.e. of our now knowing the neo-Newtonian formulation. Recall the remarks in Section \ref{symmthy} that, as I put it, `Newton's and Clarke's views are tenable'; and the remarks in this Section's preamble, about a duality being non-obvious at a certain historical epoch. Thus, interpreting the duality with `Newton's and Clarke's views', or returning to that earlier epoch: the duality illustrates (Contr).

Agreed: these `temptations', as I have called them, hint at two other important functions of duality. Namely, to prompt us to guess: either \\
\indent (i): the bare theory, the `common core', when we have no formulation of it (or only a defective, or intractable, or unperspicuous formulation), or \\
\indent (ii) another  theory `behind the duals', of which the two duals are---not representations ({\em a la} the Schema) but---approximations, in two different parts of some parameter space (different  regimes or sectors). \\
Typically, a theory (ii) will be `more different' from the duals than is the bare theory (i),  because it aims to describe more phenomena than they do: e.g. by including phenomena at higher energies, while each of the duals describes a low energy regime. (Or replace energy by another important scale, such as velocity or curvature.) Thus for Example (1), with the bare theory (i) taken as a neo-Newtonian formulation, a putative theory (ii) could  be general relativity. Clearly, guessing such theories, when one only knows the two duals, is hard!  

Obviously, these two heuristic functions of a duality are scientifically very important: and I will briefly return to them in Section \ref{string}. But for most of this paper, I can set them aside. As does much of the philosophical literature about dualities: for it concentrates on dualities' connections with issues in logic and ontology, such as theoretical  equivalence (as this paper does) and symmetry, rather than issues in methodology.\footnote{\label{heuristic}{Exceptions include: Dawid (2013; 2017 Sections 4.2, 4.3), Rickles (2011: 66; 2013: 319; 2013a, 67-78; 2017: 66), De Haro's discussion of (i) and (ii) for gauge/gravity duality (2016 Section 2.1; 2016a, Section 2.1; 2017, Section 3.2), and his general discussion of the heuristic function (2018). The latter distinguishes (i) and (ii): De Haro calls (i) a `theoretical function' (especially Section 4.1.2), and  calls only (ii) a `heuristic function', and its theory  a `successor theory' (Sections 4.1.3 and 5).}}   \\

Apart from illustrating (Contr), this Example also has four noteworthy Features. We will see them recur, or be varied, in other Examples. And Features 1) and 2) will bear on whether an Example illustrates (Contr) or (Diff):---\\
\indent 1): {\em Toy cosmology?} It is natural to think of each dual (and the bare theory) as a `toy cosmology': not least because of the historical influence of  point-particle world-pictures,  as advocated by such figures as Boscovitch and Laplace. That is: It is natural---or at least, it was traditional---to take each formulation as saying that (a) there are no bodies apart from the $N$ particles (i.e. there is no environment of the system---which is the cosmos!) and that (b) there are no forces apart from the given Galilean-invariant one, such as Newtonian gravitation. \\
\indent 2): {\em Complete set of states?} In each dual, we have what the tradition, the textbook, would regard as the complete set of states for the `toy cosmos', that are compatible with the specification of absolute rest that has been made.\\
\indent 3): {\em The size of the duality group?} In this Example, there is a whole Ôboost groupÕ worth of duals, i.e. $\mathR^3$-worth. For any vector in $\mathR^3$ can represent the velocity with which to boost from a fiducial specification of the absolute rest. But we focus on just two duals.  (In the jargon of {\em duality groups}: we take the duality group to be $Z_2$.) \\
\indent 4): {\em More variety} Of course, Newtonian point-particle mechanics, as usually understood, encompasses varying $N$; while we have implicitly taken a fixed value of $N$. But this was just to keep matters simple. That is: the simplest strategy is to have each bare theory (and so its models) fix $N$, and to express the intended generality of the original theory by simply universally quantifying over $N$. But maybe, one could formulate a more abstract bare theory that encompasses all $N$.\\ 

Note also that one could write down several  Examples that are `cousins' of Example (1). Thus one could write down theories other than point-particle mechanics, e.g. fluid mechanics, with two contrary specifications of absolute rest. The natural choice for the bare theory would again be a formulation of e.g. fluid mechanics using a neo-Newtonian spacetime. And the two duals would be formulations set in a Newtonian spacetime, that disagree with each other about absolute rest. These duals will again illustrate (Contr). And broadly speaking, the discussion in Example (1) about the state space, the definition of the duality maps, and the four noteworthy Features, will carry over, {\em mutatis mutandis}.\footnote{I say `broadly speaking' because there will be some substantial differences. For a start, the configuration space of a fluid is infinite-dimensional. So the technical details of the states, quantities and duality maps will be much more intricate; and in Feature 4), we will be universally quantifying, not over an integer $N$ representing the number of particles, but over e.g. the real number $\rho$ representing the density of an incompressible fluid. Besides, Feature 1) falters, in that we do not normally think of a sample of a classical fluid as the entire material content of the universe: but agreed, this may just reflect the historical contingency that no such world-picture was as influential as its point-particle `rivals' by such figures as Boscovitch and Laplace.}  

Similarly, one could write down theories, of mechanics or of electromagnetism, `in the spirit of Lorentz' (i.e. before he acceded to the Einsteinian abandonment of absolute simultaneity). Roughly speaking: in each formulation, a light cone structure is added to $\mathR^4$: and thus, applying the Einsteinian definition	of simultaneity, an operational notion of simultaneity relative to each inertial observer (timelike congruence) is added.   But the duals (models) are to contradict each other in their specification of the `invisible', `underlying' absolute simultaneity relation. So these duals will again illustrate (Contr). In these `cousin'  Examples, one would no doubt take the bare theory to use Minkowski geometry. And broadly speaking, the above discussion about the state space, the duality maps, and the four noteworthy Features, will again carry over, {\em mutatis mutandis}.  In particular, as in Example (1): the temptation to say that the contrary specifications of absolute simultaneity are `gauge', or `a distinction without a difference' etc., should be set aside. For  Lorentz's views ca. 1905-1912, before he acceded to Einstein's vision, were entirely tenable.

\subsection{(2):  Position-momentum duality in elementary quantum mechanics}\label{Ex2}
I turn to the unitary equivalence, by Fourier  transformation, between the position (i.e. Schr\"{o}dinger) representation and the momentum representation in elementary wave mechanics. For simplicity, we consider a single spinless non-relativistic particle in one spatial dimension. Thus one might take the bare theory to be the Hilbert space $L^2(\mathR)$, equipped with its algebra of (say: bounded) linear operators. Then the idea is: the left dual is fixed by the choice of position, the right by the choice of momentum. \\
 
 It is usual to think of these choices as just `choices of basis'. More precisely, so as to respect the fact that position and momentum are continuous quantities, and so do not have eigenvectors: as `choices of a spectral family of projectors'. The unitary Fourier  transformation is then a passive transformation sending an arbitrary physical state, expressed in position representation, to the momentum representation of the same physical state: and so preserving all quantities' expectation values. A bit more precisely, using an elementary wave-mechanical formalism: a wave function $\psi: \mathR \rightarrow \mathC$ is mapped to its Fourier transform $\tilde{\psi}: \mathR \rightarrow \mathC$. This is of course not the identity map on functions: a sharply peaked function is mapped to a spread-out one, and {\em vice versa}, attesting to the uncertainty principle. But we interpret it as the identity function on physical states, since we define a corresponding map on physical quantities (represented as linear operators) in such a way that any quantity's expectation value is preserved.  Namely, one shows that:\\
 \indent  (i) the Fourier transform $\psi \mapsto \tilde{\psi}$ is a unitary map $F$ (for `Fourier') on the state-space; so that \\
 \indent  (ii) the usual (Schr\"{o}dinger) representations of position and momentum as $Q: \psi(x) \mapsto x\psi(x)$ and $P: \psi(x) \mapsto -i \frac{d}{dx} \psi(x)$  respectively, are related by: $P = F^{-1}QF$; so that \\
 \indent  (iii) the preservation of any expectation value between the `left' and the `right' amounts, essentially, to  an inner product being the same when calculated in two different orthobases, related by a unitary map. (For details, cf.: at a pedagogic level, e.g. Jordan (1969: Section 18); or more rigorously, Prugovecki (1981: Sections 4.3, 4.5), Takhtajan (2008: 89-92).)
 
In terms of our notations, $d_s$ is the identity map on physical states, and $d_q$ is the identity map on physical quantities.  So of course  eq. \ref{obv1}, thus interpreted, is satisfied. But neither map is the identity map on states', or quantities', expressions as, respectively, complex functions on $\mathR$, and linear operators on such functions. Interpreting eq. \ref{obv1} in terms of these expressions, it expresses the equality of an inner product  calculated in unitarily related orthobases. 

But this usual way of thinking of the duality does {\em not} serve my present purpose, viz. exemplifying the Remark. For it clearly does not make the left and right duals disagree, in either of our senses (Contr) or (Diff). The two duals' claims {\em do} agree, though they are expressed in different languages (`bases').\footnote{Thus it needed no  fancy footwork in order to satisfy eq. \ref{obv1}: unlike Example (1), with its two options for defining $d_s$ and $d_q$, that carefully made only one of the two maps the identity on physical states or quantities (cf. the start of Section \ref{Ex1}).

On the other hand: this usual way of thinking of the duality is an important prototype for  other dualities where the claims of the left and right duals {\em agree}. Example (4) will be another example from quantum theory, where the duality map is a unitary  transformation that we think of as `passive' and `preserving all the physics'. And {\em bosonization} is another (much more advanced) quantum example, again with a unitary equivalence. Here, a quantum field system in 1+1 dimensions can be `written down' with bosonic excitations, and also written down with fermionic excitations: with the same degrees of freedom, or subject-matter, in both formulations (2017, Sections 4,5). In this paper, Example (5) will be an elementary {\em classical}  example of duals that agree.} \\ 

However, we {\em can} get an illustration of (Diff) by thinking of this duality differently. We need `to think like Bohr': that is, to take the position and momentum `perspectives' (`windows', `contexts') to exclude each other. Where `exclude' means---not so much `contradict', as---`cannot be adopted together'. 

Thus the left dual is to be {\em only} about position. We can take its set of states to be just the set of probability distributions for position; so in an elementary wave-mechanical formalism, a state is a probability density given as the squared modulus of the wave-function: $p(x) \equiv | \psi (x) |^2$. And the left dual's quantities are to be just position $Q$ (i.e. multiplication by the argument-variable $x$) and the (Borel) functions of position, $f(Q)$: their expectation values are then, in the elementary formalism, $\langle f(Q) \rangle = \int  dx \; f(x) p(x)$. Similarly, the right dual is only about momentum. Its states are just the probability distributions for momentum, $p(k)$, where we  write the argument variable as $k$ to signal the interpretation as momentum; and its quantities are just the (Borel) functions of momentum, $f(P)$, with expectation values  $\langle f(P) \rangle = \int  dk \; f(k) p(k)$. (We can put these choices in the slightly more abstract language of algebras of bounded operators. On the left, we can take the algebra of quantities to be just position, together with functions of position: more precisely,  the abelian von Neumann algebra generated by position. And similarly on the right: we can take  the abelian von Neumann algebra generated by momentum.) 

We now define the duality map $d_s$ to map the state $p(x)$ to the {\em same} mathematical real function: but now interpreted as a probability density for momentum (so that we again write the argument variable as $k$). So the  mathematical duality map $d_s$  is the identity map on probability distributions on $\mathR$: but as interpreted on physical states, it is  an active transformation---{\em not} the identity map. Similarly for quantities: each Borel function $f$ is mapped to itself, as a pure mathematical function, but the interpretation changes: $d_q: f(Q) \mapsto f(P)$. (Again, in the language of bounded operators: the duality map $d_q$ on quantities is the identity map on linear operators: but interpreted as an active transformation, the non-identity map, on physical quantities.)
 
With these definitions, the `preservation of values'  in our Schema, eq. \ref{obv1}, is
{\em not} a preservation of physical, interpreted expectation values, as in the usual way of thinking of the duality: (again, the motto is illustrated). Rather, $d_s$ and $d_q$  preserve the `shape' or `graph' of mathematical real functions $p$ (a probability density), and $f$ (a Borel function).\\

It is clear that this gives a case of (Diff). For the duals describe different subject-matters: position and momentum respectively.\footnote{ {\label{noworries}} {We noted towards the end of Section \ref{s-matters} that in a quantum theory, non-commutation made the idea of different subject-matters more subtle than the classical construal as cross-cutting partitions. But as announced there, this subtlety does not affect this paper: and we set it aside. Indeed Feature 1), just below, notes that the duals might be used to study different spinless quantum particles on distinct lines---making the different subject-matters, i.e. spectral families, an utterly unproblematic case of orthogonality in Lewis' sense.}}  

It is also clear that (unlike Example (1)'s case of (Contr)) there is {\em no} temptation---even in hindsight, when one has long been accustomed to the duality--- to say that there is no real difference between the duals, or that the contrast between position and momentum is `gauge', or `a distinction without a difference', or `a sign that we should move to a mechanics that eliminates the contrast'.  \\  

Let us compare this with Example (1), as regards the four noteworthy Features. For the first three, we get the opposite verdict to that in Example (1); for the fourth, the Examples are similar:---\\
\indent 1): {\em Toy cosmology?} It is not natural to think of each dual (nor the bare theory) as a toy cosmology. There might well be an environment to the system. In particular, there could be another replica system, another spinless non-relativistic quantum particle confined to another (non-intersecting) spatial line. And we could study our original given system via its position probability distributions, and the replica system via its momentum distributions. If so, the two duals would `describe different subject-matters' in two senses of that phrase: both different quantities, and different systems. Cf. footnote \ref{noworries}.  \\
\indent 2): {\em Complete set of states?} In each dual,  we {\em do not} have what the tradition, the textbook, would  regard as the complete set of states for the system. There is only position information on the left, and only the momentum information on the right. On each side the `other information', contained in the phase of the wave-function, has been erased by taking the modulus squared.   
(Agreed: we do have, on the left, all the probability distributions on the real line. So we have on the left all possible `states' in a relative sense, i.e. relative to the specification we have made of Ôwhich quantityÕ. Similarly on the right. Thus relativized, this is like Example (1).) \\
\indent 3): {\em The size of the duality group?} Given the left dual, defined as say the spectral family for position, and the idea of a unitary  transformation from it to a right dual, there is a vastly richer set of definable right duals than just the $\mathR^3$-worth we saw in Example (1) (corresponding to the various possible boost velocities $v \in \mathR^3$). After all, the unitary  transformations on $L^2(\mathR)$ are an infinite-dimensional set. (Agreed, there are various ways to drastically cut down the number of duals: by strengthening the definition of a dual from `a spectral family with $\mathR$ as spectrum' to, e.g., the abelian von Neumann algebra such a family generates; and by constraining the unitary  transformation considered.)\\
\indent 4) {\em More variety}: Again, the idea of the duality encompasses a generality I have not yet mentioned. The quantum particle that is the system can be in  two or three spatial dimensions, instead of one; and there can be 1, or 2, or .... $N$ particles. Again, the simplest strategy is to keep the bare theory reasonably concrete (hence our choice of $L^2(\mathR)$), and to express the intended generality  by simply universally quantifying, i.e. having many close-cousin dualities. \\

\subsection{(3):  Kramers-Wannier duality in classical statistical mechanics}\label{Ex3}
It is interesting to note that this famous duality  is conceptually similar to Example (2): at least, `conceptually similar' if we adopt the broad-brush, or bird's-eye, perspective of a philosopher! Though this duality is unfamiliar to philosophers, it is worth briefly sketching: both for this reason and because---like Examples (1) and (2)---it will help substantiate the Remark. (I will follow Thompson (1972: Section 6-2). For more details, including many generalizations, cf. e.g. Savit (1980), Lavis and Bell (1999).)

We can take the bare theory  to be the classical equilibrium statistical mechanics of a two-dimensional square lattice with the Ising Hamiltonian: more specifically, the canonical ensemble with the Boltzmann probabilistic weights for a configuration $s$ given by $\exp (-\beta H[s])$, where $H$ is the Ising Hamiltonian and $\beta \equiv 1 / kT$ is the inverse temperature. We take the duals, i.e. the models of the bare theory, to be approximations (viz. expansions involving a sum of exponentials) to the partition function $Z \equiv \Sigma_s \;  \exp (-\beta H[s])$ that are valid at low and high  temperatures $T$, respectively: say, low  on the `left' and high on the `right'. 

So in this Example, `model' does not mean `representation' in the sense of representation theory, as it usually does for our Schema. Here instead, our use of `model'  sits happily with the word's usual connotation that a model is an approximation; (so  the warning (ii), at the start of Section \ref{schema}, can be ignored here---though not in any of this paper's other Examples). But to give a more precise definition of the duals, i.e. the models, it will be clearest to first sketch the idea of the duality.  

It turns out that, since a square lattice is self-dual in a certain geometric sense,\footnote{That word again! But I need not define it: it is a matter of the lattice being mapped to (as isomorphic copy of) itself under, being  a fixed point of, a certain geometric construction.} the low and high  temperature expansions are very simply related. If we write the partition function in terms of the dimensionless inverse temperature $\nu := J/kT$, where $J$ is the nearest-neighbour coupling  in the Hamiltonian, then the low temperature (large $\nu$) expansion  $Z(\nu)$ obeys (with $N$ the number of lattice sites)
\be 
Z(\nu)= Z(\tanh^{-1}(\exp (-2 \nu)))2^{1-N}(2 \sinh 2 \nu)^N \; .
\ee
Now we define $\nu^*$ by $\tanh \nu^* := \exp (-2 \nu)$, so that $\nu^* = 0/ \infty$ iff $\nu = \infty / 0$ respectively. So low temperature i.e. large $\nu$ corresponds to a high conjugate temperature i.e. small $\nu^*$. Then we can write: 
\be 
\label{kwlink}
Z(\nu)= Z(\nu^*)2^{1-N}(2 \sinh 2 \nu)^N \; .
\ee
relating the expansions for low and high  temperatures. 

Historically, the original scientific importance of this duality lies in its implication that singularities of the partition function, representing phase transitions, must occur in pairs. For this implies that if there is only one singularity, it must occur at a critical temperature given by $\nu^* = \nu$, i.e. when $\tanh \nu = \exp (-2 \nu)$ i.e. $\sinh 2 \nu = 1$ or $\nu_c := \frac{1}{2} \sinh^{-1}(1)$. In other words, the critical temperature $T_c \equiv J/k \nu_c = 2J / k \sinh^{-1}(1)$. By this kind of reasoning, Kramers and Wannier deduced in 1941 that the 2-dimensional Ising model has a phase transition, and the value of the critical temperature: three years before Onsager, in 1944, solved the model exactly.\footnote{For my rationale in saying `kind of reasoning', and more references about how the duality between high and low temperature regimes can be generalized to other models, cf. (briefly) Thompson (1972: 117, 154); and for details, Savit (1980), Lavis and Bell (1999: Sections 8.2-8.3, 206-211). For the history of the Ising model, cf. Brush (1967).}\\

So much by way of sketching the duality. To put it in terms of our Schema: as we said, we take the bare theory  to be the exact canonical ensemble of the lattice at all temperatures: more precisely, the family (parameterized by varying $T$) of all the exact partition functions. So we here set aside the microscopic configurations ($s$ in $\exp (-\beta H[s])$ above), and take a state in our Schema ($s$ in Section \ref{schema}'s notation) to be a partition function $Z(T)$, with the values of  quantities such as free energy extracted by taking logarithms, differentiation etc. in the usual way. The two duals are then the approximations, i.e. the expansions, valid at low and high  temperatures. We do not need their exact form. We only need to note that: taking  the left dual as the family of expansions parameterized by $T$ being in some low range $[T_1, T_2] \subset \mathR$, i.e. by large $\nu := J/kT$ in the range $[ J/kT_2, J/kT_1 ]$, corresponds to the right dual being the expansions parameterized by small $\nu^*$ in the range $[\tanh(\exp ( J/kT_1)), \tanh(\exp ( J/kT_2))]$. Then: eq. \ref{kwlink} implies that  the duality map on states $d_s$ is defined by 
\be 
\label{kwds}
d_s: Z(\nu) \mapsto Z(\nu^*) := \frac{Z(\nu)}{2^{1-N}(2 \sinh 2 \nu)^N} \; .
\ee \\

It is clear that this is similar to Example (2) in several ways. For us, the main point is that it illustrates the Remark, by being a case of (Diff). The left dual is about the low temperature regime,  the right dual about the high temperature regime---different though isomorphic subject-matters. This is similar to Example (2)'s left dual, the position description (`perspective',`window') being isomorphic to the momentum description. Indeed, it illustrates (Diff)'s idea of cross-cutting partitions more simply than Example (2) did, since it is unclouded by quantum theory's non-commutation of quantities: cf. footnote \ref{noworries}, and Feature 1) below.

And as in Example (2): there is {\em no} temptation---even in hindsight, when one has long been accustomed to the duality---to say that there is no real difference between the duals, or that the contrast between low and high temperature is `gauge', or `a distinction without a difference', or `a sign that we should move to a statistical mechanics that eliminates the contrast'.  \\

The noteworthy Features yield further similarities. In short: The first two Features are similar to Example (2); the third Feature is more like Example (1) (but more `extreme'); and the fourth Feature matches both Example (1) and Example (2) (which agreed with each other). In detail:---\\
\indent 1): {\em Toy cosmology?} It is not natural to think of each dual (nor the bare theory) as a toy cosmology. Of course, there might well be an environment to the system. In particular, there could be another replica system, another Ising lattice. And we could study our original given system at low temperature, and the replica system at high temperature. If so, the two duals would `describe different subject-matters' in two senses: both different temperature ranges, and different systems. And thanks to considering different systems, this gives another case of subject-matters that are orthogonal  in Lewis' sense. In terms of partitions and worlds: a cell for the left dual describing a lattice at a low temperature $T$ intersects all cells for the right dual describing a lattice at a high temperature---since worlds in these intersections contain two lattices, one at the low temperature $T$ and the other at some high temperature---which temperature depending on which cell of the right dual we consider. \\
\indent 2): {\em Complete set of states?} In each dual,  we {\em do not} have what the tradition, the textbook, would  regard as the complete set of states for the system. For first, the duals are approximations; and second, there is only low temperature information on the left, and only high temperature information on the right.  (Besides, as noted above: since this is thermal physics, we have in each dual set aside the microscopic conception of state, i.e. all the microscopic configurations of the lattice.) \\
\indent 3): {\em The size of the duality group?} On our definitions, the left dual state at inverse temperature $\nu \equiv J/kT$ has a unique right dual at inverse temperature $\nu^*$.  (The real-parameter family of states is the dual `as a whole'.) So far as I know, there is no `knob' in each dual's specific structure, analogous to Example (1)'s choice of absolute rest or Example (2)'s choice of unitary  transformation, that can be `twiddled' to yield, for a given left dual, a whole family of right duals, analogous to Example (1)'s $\mathR^3$-worth of boosted duals.\\
\indent 4): {\em More variety}: Again, there is a generality I have not yet mentioned: though it is more complex and varied than just changing the value of a parameter or two, such as $N$, the number of lattice-sites (or the number of particles in Examples (1) and (2)), and $J$ the coupling. For it turns out that the duality generalizes, and not just to Ising models in other dimensions, but to many other models: both to statistical systems and to field theories, pairing weak and strong couplings.   Savit (1980) is a thorough review.

\subsection{(4):  The unitary equivalence between Heisenberg and Schr\"{o}dinger pictures in quantum mechanics}\label{Ex4}
This Example is like Example (2), in two ways. It returns us to elementary quantum mechanics. And, like the usual way of thinking of the Fourier transform between position and momentum, the duality map is a unitary  transformation that we think of as  a `passive'  transformation that `preserves all the physics'. 

This second point means that it does not serve our present purpose, viz. exemplifying the Remark: since the left and right duals do not disagree, in either of our senses (Contr) or (Diff). So I will treat it only briefly.\footnote{The main merits of including it are that: (i) on reading the Schema, it naturally comes to mind as a putative example; and (ii) it is unlike the previous Examples in being all about dynamics: which  has been downplayed hitherto (as I announced), and which was wholly absent from the equilibrium theory of Example (3).}   

 To make this Example neat, one has to suppress Section \ref{details}'s requirement that a duality should respect the dynamics (whether in Heisenberg or Schr\"{o}dinger  picture). Besides, the one-parameter group of unitary operators that carries one between the pictures depends on the system, i.e. the Hamiltonian, concerned. So to make the Example neat, it is best to make the bare theory so specific as to choose a Hamiltonian. Then the simplest tactic is to define the bare theory to consist simply of all appropriate (e.g. continuous) temporal sequences of assignments of values to all desired quantities---which are then represented in Hilbert spaces by the Heisenberg and Schr\"{o}dinger pictures (duals).\\
 
  I will not linger to put all this in notation, but will just turn to the noteworthy Features.
 
 In short: The first Feature is similar to Examples (2) and (3), and unlike (1); the second Feature is the other way round---similar to Example (1) and unlike Examples (2) and (3); the third Feature is  like Examples (1) and (2); and the fourth Feature matches all three previous Examples (which agreed with each other). In detail:---\\
\indent 1): {\em Toy cosmology?} It is not natural to think of each dual (nor the bare theory) as a toy cosmology. In general, there are of course  other systems, and some of these might be replicas of the given one (with the given Hamiltonian). \\
\indent 2): {\em Complete set of states?} In each dual,  we have what we would  regard as the complete set of states for the system. \\
\indent 3): {\em The size of the duality group?} There is a `plenitude of duals'. As we learn from textbook discussions of the interaction picture, one can `divide the labour' of time-evolution between states and quantities in different ways, additional to the `only states' and `only quantities' of the  Schr\"{o}dinger and Heisenberg pictures. In principle, each expression of the Hamiltonian as a sum defines such a `division of labour'. \\
\indent 4): {\em More variety}: Again, there is a generality I have not yet mentioned: though it is more complex and varied than just changing the value of a parameter or two. Namely: the duality encompasses varying Hamiltonians. So here is an analogy with previous Examples, especially (3). And again, the simplest tactic to express the intended generality is to just  universally quantify over Hamiltonians.

\subsection{(5):  The Legendre  transformation between Lagrangian and Hamiltonian classical mechanics}\label{Ex5}
My last elementary Example is the Legendre  transformation between Lagrangian and Hamiltonian classical mechanics. For my purposes, it is in broad terms like Example (4). For (i): on reading the Schema, it naturally comes to mind as a putative example; (ii) unlike  Examples (1)-(3), it is about dynamics; and (iii) most important for us: it does not serve our present purpose, viz. exemplifying the Remark: since the left and right duals do not disagree, in either of our senses (Contr) or (Diff). 

Agreed: a Lagrangian and a Hamiltonian description of, say, $N$ point-particles, or of a spinning top, differ in some of their advantages and disadvantages, both formal and interpretative. (My own effort to articulate some of these advantages and disadvantages is Butterfield (2004, 2006).) But surely, no one would say that the descriptions contradict each other, {\em a la} (Contr), or that they have different subject-matters {\em a la} (Diff).  So like Example (4), I will treat it only very briefly, just sketching the idea.\\

Let us for simplicity confine ourselves to finite-dimensional systems: and also fix the configuration space $Q$ once and for all. Then the state-space for the Lagrangian dual (model, in our sense)  is the tangent bundle  $TQ$, and the state-space of the Hamiltonian dual (model) is the cotangent bundle $T^*Q$. 
  The idea is that the duality map $d_s$ on states should {\em be} the Legendre transformation (which is also called: `fibre derivative'). It should map the Lagrangian `physical state' $(q, {\dot q}) \in TQ$ to the Hamiltonian `physical state' $(q, p) \in T^*Q$. For this to be work as neatly as possible, various conditions must be satisfied. The two main ones  are:\\
\indent \indent (a) Since we must have $p = {\partial L} /{\partial {\dot q}}$, we also require that the Lagrangian dual be equipped with the Lagrangian $L$, as a scalar function on $TQ$: $L: TQ \rightarrow \mathR$; and that the Hamiltonian dual be equipped with the Hamiltonian $H$ as a scalar: $H: T^*Q \rightarrow \mathR$. (So this implies a time-independent dynamics on both sides.) There is then enough structure in the left Lagrangian dual for the map $d_s$ to be defined; and similarly for the inverse map $d^{-1}_s$.\\
\indent \indent (b) $L$ must be such that the Legendre transformation $d_s: (TQ, L) \rightarrow (T^*Q, H)$ is a diffeomorphism: its inverse is then $d^{-1}_s: (T^*Q, H) \rightarrow (TQ, L)$. This is called the Lagrangian and Hamiltonian being {\em hyperregular}.\\
There is a wealth of deep and beautiful theory hereabouts. But it is not for this paper.\footnote{A classic exposition is Abraham and Marsden (1978: especially theorems 3.6.2, 3.6.8, 3.6.9). Recent uses  of this Example in connection in philosophical discussions about how best to define theoretical  equivalence include Barrett (2015, 2017), Teh and Tzementzis (2017).}  \\    

 I turn to the noteworthy Features.
 
In short: the Features are like Example (4)---perhaps unsurprisingly, since both Examples are about two `total frameworks' for describing dynamics. Thus the first Feature is similar to Examples (2) (3) and (4), and unlike (1); the second Feature is similar to Examples (1) and (4) and unlike Examples (2) and (3); the third Feature is  like Examples (1), (2) and (4); and the fourth Feature matches all four previous Examples (which agreed with each other). In detail:---\\
\indent 1): {\em Toy cosmology?} It is not natural to think of each dual, which  uses a fixed finite-dimensional configuration space $Q$,  as a toy cosmology. In general, there are countless other systems; and some of these might be replicas of the given one (with the given Lagrangian/Hamiltonian). Agreed: This contrast with Example (1) reflects history and habit, as I mentioned under Feature 1) in Section \ref{Ex1}.  \\
\indent 2): {\em Complete set of states?} In each dual,  we have what we would  regard as the complete set of states for the system. \\
\indent 3): {\em The size of the duality group?} There is a `plenitude of duals'. As we learn from textbook discussions of Routhians (e.g. Goldstein et al. 2002: 348), we can `pick and choose' about how to treat the time derivatives of the configurational variables $q_1, ..., q_N$. That is: we can adopt a mixed Routhian formalism using velocities for the first $M$ variables ${\dot q}_1, ... , {\dot q}_M$ and canonical momenta for the remainder $p_{M+1},..., p_N$. \\
\indent 4): {\em More variety}: Again, there is a generality I have not yet mentioned: and again, it is more complex and varied than just changing the value of a parameter or two. And this is so, even if we set aside the adaptation of the Lagrangian and Hamiltonian frameworks, beyond finite-dimensional mechanics, in e.g. continuum mechanics and electromagnetism. Namely: within finite-dimensional mechanics, there is the great variety of configuration manifolds $Q$; and for a fixed $Q$, all the various Lagrangian/Hamiltonian scalar fields on the tangent bundle/cotangent bundle. So here is an analogy with previous Examples, especially (3) and (4). And again, the simplest tactic to express the intended generality is to just  universally quantify over manifolds and Hamiltonians.  \\

{\bf To sum up this Section}: With five examples, we have illustrated the Schema for duality, and established my main Remark. Namely:  duals can disagree with each other, either by:\\
\indent \indent (Contr): making contrary assertions about a common subject-matter, such as the standard of absolute rest: as in Example (1); or by :\\
\indent \indent (Diff): making claims about distinct but suitably `isomorphic' subject-matters: such as the continuous quantities, position and momentum, in elementary quantum mechanics (Example (2)); or (approximations to) the low and high temperature regimes of an Ising model  (Example (3)).\\

So as announced in Section \ref{prosp}, my tasks now are:\\
\indent \indent (i)  to spell out the Implication of this Remark for proposed definitions, in logic and philosophy, of theories being `equivalent' (Section \ref{logic});\\
\indent \indent (ii) to argue that some string-theoretic dualities also  illustrate the Remark (Section \ref{string}).


\section{An Implication about theoretical  equivalence}\label{logic}
The Remark has a simple implication---I shall say `Implication'---for accounts of theoretical equivalence. Namely: two dual theories  satisfying (Contr) or (Diff) might get formalized so as to be logically equivalent---where I use `formalized' and `logically equivalent' with their standard meanings, in logic and philosophy. But obviously such duals are not equivalent---as `equivalent' is normally understood. So such cases show that logical equivalence is too weak an explication of theoretical equivalence. And so also, therefore, is any weakening of logical equivalence. 

As I explained at the end of Section \ref{prosp}, this Implication, though simple,  is worth stressing. For the recent philosophical literature on theoretical equivalence has focussed on various weakenings of logical equivalence: a focus motivated by logical equivalence  being clearly, in one regard, too strong. So it is worth seeing that---notwithstanding the literature's rationale for weakening  logical equivalence---there is also a problem `in the other direction'. That is: a defect of  being too weak, rather than too strong (as an explication of theoretical equivalence).

I will fill out this Implication in three Subsections. The first is a warning about jargon, especially `theoretical equivalence' and `physical equivalence' (Section \ref{jargon}: which returns us to Section \ref{intsemics}'s advertisement for intensional semantics). The second spells out the Implication for logical equivalence (Section \ref{implic}). The third shows that the Implication also holds for notions weaker than logical equivalence, which have been proposed in the literature (Section \ref{implic2}).


\subsection{A warning about jargon}\label{jargon}
We saw at the end of Section \ref{prosp} that `theoretical equivalence' is a term of art. While all agree that we need some notion of `making the same claims about the world', there is no agreed definition.  (This is like `proposition' in the philosophy of logic and language: all agree we need some notion of `meaning of a sentence', but the exact definition varies or is lacking.) We also saw that logical equivalence, as a candidate definition, is both:\\
\indent (i): too strong, because of cases of synonymy (like `electric' and `\'{e}lectrique'): prompting the effort to give a weaker definition; and \\
\indent (ii): too weak, because of cases of homonymy (like `bank' in finance and beside a river): i.e. cases where logical equivalents disagree in their claims about the world---which will be illustrated by my Remark's Implication.

Agreed,  you might want to reserve  the phrase `theoretical equivalence' for some weakening of  the familiar logical equivalence. For the effort mentioned in (i) finds several salient notions, and there are too few good but distinctive words for them. Or you might want, regardless of the hunt for the best weakening of  logical equivalence, to reserve  `theoretical equivalence' for some formal notion that sets aside issues of interpretation, simply because the adjective `theoretical' connotes a contrast with `interpretation'. So, agreed: if you do this, the pairs of theories the Remark draws attention to, may well get classified as theoretically equivalent. (`May well' because it will depend on your exact definition of `theoretical equivalence'.) But of course, the implication of the Remark  remains. Namely, expressed without the jargon of `theoretical equivalence': as interpreted, two such dual theories do not make the very same claims about the world.

And to reflect the Remark, it will of course be natural (to avoid confusion) to introduce another phrase  for `making the very same claims about the world'. As mentioned in Section \ref{others}, De Haro (2016, Section 1.2; 2016a, Section 1.2) suggests {\em physical equivalence}: a phrase adopted by our (2017: Sections 1.1, 3.2.3(1), and also  by others, e.g. Huggett (2017: especially 85-86). 

Fair enough, but {\em beware}: this phrase can also be misleading. For in philosophy, we often distinguish the vast plethora of all (logically and-or metaphysically) possible worlds, $W$, from its subset, the set $N$ of all the nomically possible worlds: the worlds that obey the actual laws of nature. (At least: we right-thinking Humeans, believing that the laws of nature are contingent, do so.) And the nomically possible worlds are often called the `physically possible' worlds, in a nod to the success of physics in formulating our best guesses about the laws of nature. Thus, with {\em this} jargon in mind, one naturally hears `physical equivalence'  as meaning `equivalence across all the physically possible worlds', i.e. across $N$: which, since $N \subset W$, will be a {\em weaker} relation than `equivalence across all the possible worlds', i.e. across $W$. This means that the label `physical equivalence', for our notion of `making the very same claims about the world', can be misleading, since my point---the Implication---is that some pairs of theories, when interpreted, make different claims and yet are logically equivalent.

And finally: another confusing feature of the various jargons is  that the logic books' `logical equivalence' is not really the same as `equivalence, i.e. sameness of truth-value, across all the worlds': even when we take $W$ as the set of all logically  possible worlds, so that we naturally hear `logical equivalence' as equivalence across all of $W$. For again: the logic books' `logical equivalence' is formal---tied to a vocabulary, and disregarding any intended meanings---and so will rule e.g. `The weather is fine' and `Il fait beau' as inequivalent. 

This last point returns us, of course, to Section \ref{intsemics}'s advertisement for intensional semantics. {\em That} is a framework for analysing meaning etc. that respects intended meanings: and according to which, `same meaning', i.e. our desired notion of `making the  same claim about the world', indeed corresponds to `equivalence across all the worlds'. More power to its elbow ...

\subsection{The Implication: for logical equivalence}\label{implic}
The Remark---dual theories can disagree with each other---implies a limitation of standard approaches for formalising physical theories; (indeed, as we shall see: theories about any topic). For two such disagreeing dual theories might get formalized so as to be equivalent `within the meaning of the act', i.e. equivalent according to the approach's precise definition of `equivalence'. Or to put it more charitably, since `theoretical equivalence' is, after all, a term of art (Section \ref{jargon}): the Remark implies that these approaches' use of `theoretical equivalence' has a misleading connotation. 

This Implication follows because the duality implies that for this mishap, this misleading formal equivalence, to occur, the only thing needed is appropriate homonymies. That is: the same word being  used in the two theories, but with mutually `inverted' meanings---meanings that of course formalization sets aside. 

Recall Section \ref{idea}'s warning about the physics jargon `dictionary': so now the dictionary will pair the same word with mutually inverted meanings, rather than different words with the same meaning, as `dictionary' usually connotes.  (But not all the words used in the theories need to have their meanings `inverted' by the dictionary: just an appropriately chosen subset.) 

I shall first explain this Implication, for the most well-known approach to defining `theoretical equivalence': namely, logic books' notion of  logical equivalence. The next Subsection explains how the Implication applies to other approaches.\\

Thus the logic books take theories as sets of  sentences (closed under logical deduction); and then define theories  as {\em logically equivalent} if they are stated in the same language (i.e. have the same non-logical vocabulary: also known as `signature'), and are made true by the same interpretations, i.e. have the  same set of models (in the sense of `model' used by logic and semantics, rather than {\em our} sense in Section \ref{idea}!) Cf. e.g. Hodges (1997: 37-38).  So for this approach, the Implication is that two dual theories that disagree with one another (in sense (Contr) or  (Diff)) may well get formalized so as to be logically equivalent, because their different claims about the world are `isomorphic claims'.

We can make vivid this Implication, this mishap, in terms of the proverbial Alice on the left, i.e. advocating the left dual, and Bob on the right, advocating the right dual. If Alice and Bob are `unwise enough' to use the same language in their advocacy of their doctrines---in the {\em syntactic} sense of `language'---and an appropriate subset of the words they both use have appropriately `inverted' meanings, then: what they {\em say}, in the  syntactic sense, i.e. the words they utter, could be the same, despite their theories disagreeing. In which case their theories might well get formalized, i.e. regimented in to a formal language, so as to render the resulting formal theories logically equivalent. 

Thus take Example (3), Kramers-Wannier duality (Section \ref{Ex3}). Suppose Alice speaks standard English,  so that in advocating the left dual, i.e. the expansion for the Ising lattice at low temperatures, she says `low temperature expansion' etc. And suppose the  semantics (intended interpretation) of Bob's language is `high-low' inverted with respect to Alice's, i.e. with respect to standard English. So Bob, in advocating the right dual, i.e. the expansion for the Ising lattice at high temperatures, and describing such a lattice at high temperature---i.e. what we (and Alice!), speaking standard English, call `high temperature'---says `low temperature expansion' etc. (Not all Bob's words need to have meanings `inverted' with respect to Alice's: apart from the `high-low' inversion, he can speak standard English, e.g.  saying `lattice' to mean lattice.) Then, thanks to the duality, Alice and Bob in advocating their different theories, may say the very same set of sentences. So the theories might get formalized so as to be rendered logically equivalent.  

Exactly corresponding comments apply, of course, to the Examples of absolute rest ((Contr): Section \ref{Ex1}), and position-momentum duality ((Diff): Section \ref{Ex2}). Thus for the latter, we can suppose Alice speaks standard English,  so that in advocating the left dual, her word `position' means position; while Bob uses the word `position' to mean momentum---with the result that they say the very same set of sentences. So their two doctrines might get formalized so as to be rendered logically equivalent.

We can also state the Implication in terms of the duality maps $d_s$ etc: and thereby illustrate our {\em motto} about how confusing language can be! Thus in Example (1), on the first definition of the duality maps: suppose Alice describes a state $s$ by saying `the $N$th particle is at rest'. Then Bob describes the transformed state $d_s(s)$ (i.e. boosted relative to Alice by the same amount as is his standard of rest)  by saying the same words. But they contradict one another: (Contr).

In Example (2), Alice describes a state $s$ by saying `the   particle has a Gaussian distribution for position centred at $x=5$'. Bob describes the transformed state $d_s(s)$ (i.e. the same probability distribution but for momentum instead)  by saying the same words. Their assertions differ, but are compatible: (Diff). For `position' in Bob's mouth means momentum (in Alice's mouth, and ours). And Bob could be referring to a different particle.

In Example (3), Alice describes a state $Z(\nu)$, i.e. an expansion of the partition function at low temperature, by saying the lattice has a free energy $F = F(Z(\nu))$ for a temperature $T = T(\nu)$. Bob describes the transformed state $d_s(Z(\nu)) := Z(\nu^*)$ (cf. eq. \ref{kwds}), i.e. the expansion at the conjugate high temperature, by saying the same words. Their assertions differ, but are compatible: (Diff). For `low' in Bob's mouth means high (in Alice's mouth, and ours). More exactly: the notation (`numeral') $\nu$ in Bob's mouth means the real number $\nu^* :=  \nu^*(\nu) := \tanh^{-1}(\exp (- 2 \nu))$. Similarly for the notation $T$ and $T^*$. And Bob could be referring to a different lattice. 	

So much by way of explaining the Implication, as applied to the notion of logical equivalence. There are three comments, (A) to (C), to make. \\
 
(A): {\em Permutation arguments}: Mention of homonymy, and especially of appropriately `inverted' meanings, is bound to remind philosophers of other permutation arguments, such as the inverted spectrum in the philosophy of perception (e.g. Shoemaker 1982), Putnam's argument for the indeterminacy of reference (1981: Chapter 2, Appendix; Lewis (1983a); Button and Walsh (2018: Chapter 2)), and Einstein's hole argument (connected to Putnam's argument by Rynasiewicz 1994). A rich field for comparison with this paper's project---but one for which I have no space.\\   

(B):  {\em The labour of formalization}: My saying `might well get formalized' is shorthand for some substantial labour. For by the standards of logic, physical theories are almost always presented very informally. Even formulations that by the standards of theoretical physics are rigorous are a far cry from what a logician would call a formal theory. This applies, not just to famous expositions of overarching general theories  (e.g.  of mechanics by Abraham and Marsden (1978), and of general relativity by Hawking and Ellis (1973)), but to expositions of the simpler, because more specific, dual theories that were Section \ref{weak;exples}'s Examples. Even the multi-variable calculus of Newtonian point-particle mechanics (Examples (1), (5)), let alone the functional analysis needed for rigorous quantum theory and statistical mechanics (Examples (2) to (4)), are difficult to capture adequately and perspicuously in formal languages, especially with a first-order logic. (At the end of Section \ref{implic2}, we will see  that this trouble is one motivation for recent proposals about theoretical equivalence.) 

Nor is it just a matter of difficulty; or of physicists' and logicians' different habits or interests as regards rigour. There are matters of principle. For as we learn in logic, continuity is a second-order notion. That is (roughly speaking): to express the difference between the reals and the rationals---whether every bounded set has a least upper bound---you need second-order quantification: which brings with it the incompleteness (lack of a complete proof procedure) for second-order logic.  (The usual response is to stay in first-order languages, and adopt, following Tarski, a first-order axiom scheme for continuity: e.g. Goldblatt (2004, 61f.), Andreka et al. (2012: 643 and references cited there).) 

So my saying `might well get formalized' is programmatic. But notice that the labour and obstacles just summarised have nothing to do with dualities. So it remains very plausible that two duals could get formalized so as to be logically equivalent: thanks to the isomorphism of model triples and the easy correspondence between the `knobs' (e.g. specifications of absolute rest, or of a range of tempertaure)  that yield (Contr) and (Diff).\\

(C): {\em The limits of formalism}: Agreed: in order for the mishap to occur---for the disagreeing duals to get formalized as logically equivalent---Alice and Bob have to be `unwise enough' to use the same words in their advocacy of their doctrines. And the person, Charlie say, doing the formalizing has to be `unwise enough' to disregard Alice's and Bob's different intended meanings of their common word(s). 

Thus the Implication turns on the platitude that formal methods in logic and model theory  pay no heed to the intended interpretations of words. They only `discern structure' by focussing on the mathematical consequences of assigning arbitrary references to  words (more precisely: non-logical vocabulary). Thus as I have admitted: the Remark and Implication are unsurprising. For {\em afficionados} of the literature on theoretical equivalence, they may be old hat. But I think they are still worth stressing, just because the literature seems not to say them loud and clear.

\subsection{The Implication: for weaker notions of equivalence}\label{implic2}
 Logical equivalence is of course just one notion of equivalence  between theories. This prompts the question: how does the Implication fare with other notions? In this Section, I will argue that since the other proposed notions of `theoretical equivalence' are weakenings of logical equivalence, the  Implication again holds good.
 
I begin by recalling that the logical tradition of taking a theory as a set of sentences, which leads to defining logical equivalence as above, has been rejected by many philosophers of science. They advocate instead a `semantic conception' of theories, saying that a theory is a set of models: though authors differ about what they mean by `model', and what are the merits of the semantic conception.  Obviously, space prevents a full discussion of this conception; and even of how the Implication fares with it. I will make just four comments about the Remark and Implication. It is the fourth that leads to proposed weakenings of logical equivalence, and so is the most important of the four.\\

\indent \indent \indent (1): {\em Better off?}: {\em Prima facie}, the semantic conception seems able to accommodate the Remark's cases of (Contr) and (Diff). For it promises to classify two such duals  as being `different theories', simply because  Section \ref{details} had different model triples on the two sides. So this suggests that this conception faces no `embarrassing Implication' from the Remark's cases of (Contr) and (Diff).

\indent \indent \indent (2): {\em Not so fast!}: On the other hand, our Schema allows dualities whose duals do {\em not} disagree. Consider, in quantum theory, position-momentum duality at the start of Section \ref{Ex2}, the Heisenberg/Schr\"{o}dinger pictures of time evolution (Section \ref{Ex4}) and bosonization; and in classical physics, the Legendre transformation (Section \ref{Ex5}). And more generally: the semantic conception will individuate theories far too finely if it declares two theories different whenever there are different models (or classes of models). For whatever exactly one means by `model', a model will have `specific structure' (cf. Section \ref{details}) that may not contribute to its representational role. So overall, the semantic conception will have some work to do to account for dualities---at least according to our Schema.\footnote{{\label{Bas2}} Halvorson (2012: Section 4.2, 193-197) argues in general, without regard to dualities, that the semantic conception will individuate theories far too finely if they are identified with a set of models. But van Fraassen (2014: 278-281) replies that the semantic conception never endorsed so fine an individuation of theories. Cf. footnote \ref{Bas1}.} 

\indent \indent \indent (3): {\em A false contrast?}: The point in (2), about the danger of individuating theories far too finely, recalls more general difficulties facing the semantic conception: some of which erode the distinction between it and the syntactic conception of theories  which was its target. In brief, this goes as follows.  (Cf. Halvorson and Tsementzis (2015) and Lutz (2017) for recent detailed critiques of the contrast between syntactic and semantic conceptions.) In order for the semantic conception to describe models' representational role, and free itself from `specific structure', it is drawn into associating words with the representational parts of models. But for model theory (in logic's sense!), and thus also for the syntactic conception, a model is tied to a specific non-logical vocabulary (predicates etc.) that gets assigned referents in the model's domain. As van Fraassen (1989: 366) put it: `models are defined, as in many standard logic texts, to
be partially linguistic entities, each yoked to a particular syntax'. So given any class of such models all tied to the same vocabulary, there is a well-defined set of sentences true in each model: which model theory calls the `theory' of the class. Agreed, this theory need not be an axiomatisable, or cast in a first-order language: but the syntactic conception was never committed to theories having either of those features.

\indent \indent \indent (4): {\em Weakening logical equivalence}: As I have announced: the literature on theoretical equivalence (whether adopting the syntactic conception or the semantic one) has focussed on a limitation of defining  `theoretical equivalence' as logical equivalence that goes in the  {\em opposite direction} than does our Remark. That is: there are pairs of theories that:\\
\indent \indent \indent \indent  (a) we intuitively think of as the same, but that:\\
\indent \indent \indent \indent (b) are not logically equivalent, in the above sense.\\
Informal mathematics gives many simple examples that are more precise than my mention of electromagnetism in English andin French. Examples: (i) the theory of Boolean algebras vs. the theory  of Boolean rings; (ii) the usual axioms for a group as a binary operation,  obeying certain conditions, on a set of elements, vs. the specification of a one-object category, whose arrows are then the group elements; (iii) topology as  axioms governing a notion of `open set', vs.  as axioms governing `closed set', vs. as  axioms governing `closure operation'. Such pairs of theories show logical equivalence is too strong as a criterion, or explication, of `theoretical equivalence'.

Of course, this limitation has  long been recognized by logicians and philosophers, including advocates of the syntactic conception. And so began the hunt  for a  logically weaker explication. For this hunt, early and recent references in logic are de Bouvere (1965), Visser (2006) and Andreka and Nemeti (2014); and recent references in philosophy are Barrett and Halvorson (2016, 2016a, 2017), Button and Walsh (2018: Chapter 5), Hudetz (2016, 2017, 2018) and Weatherall (2015, 2016). 

But this is not the place to review the ideas and results of this literature. I will just sketch one main idea, called {\em definitional equivalence} (also called `synonymy', as in my own jargon), and then emphasise that it makes no difference to the Remark and Implication. That is: dual theories that satisfy (Contr) or (Diff), as in the Remark, might well get formalized so as to be definitionally equivalent. It will then be clear that two other proposed weakenings of logical equivalence also make no difference to the Remark and Implication.\\

The idea of definitional equivalence is that starting from one presentation of a theory---say, Boolean algebras in terms of `meet' and `join' in a lattice---one can introduce definitions of the other presentation's notions (e.g. operations yielding a ring) and then rigorously deduce all  the claims, i.e. theorems, of the other presentation; and vice versa, starting with the other presentation. Such deductions are almost always conducted in the informal mathematical natural language of a textbook. But there is  a strong consensus in mathematics about which proofs are rigorous enough that they could be made wholly rigorous e.g. by being transcribed into a predicate calculus formulation of a standard set theory such as ZFC.\footnote{This consensus is, of course, the legacy---the achievement!---of the fifty years of effort, from the Dedekind-Weierstrass arithmetization of analysis and Frege's invention of predicate logic, to about 1930: namely, to show that all known pure mathematics could be expressed as a branch of set theory, and that set theory could  be axiomatized, and thus the proofs made rigorous and effectively checkable, in predicate logic.} 

Accordingly, definitional equivalence is made precise by mathematical logic (with its deliberately setting aside knowledge of intended meanings!) by taking the two presentations as formal theories, i.e. as sets of sentences in a formal language, closed under deduction: call them $T_1$ and $T_2$. To avoid trouble from ambiguous vocabulary items, e.g. the same predicate occurring in both languages, one assumes that the vocabularies of $T_1$ and $T_2$ are disjoint. Then $T_1$ and $T_2$ are  {\em definitionally equivalent} iff: (i) one can add to $T_1$ a definition of each vocabulary item of $T_2$, in such a way that within this augmentation of $T_1$ one can deduce all of $T_2$; and of course (ii) vice versa. Thus (i) is called: making a {\em definitional extension} of $T_1$, and showing $T_2$ to be (a sub theory of) a definitional extension of $T_1$. 
So: definitional equivalence is a matter of the two theories each being definitionally extendable so as to `encompass' the other. This is also called: their having a {\em common definitional extension}. 

It is clear, without revisiting the details of our Examples (1) to (3), that this notion of definitional equivalence makes no difference to the Remark and Implication. And for the now-familiar reason, viz. the logical framework  deliberately sets aside intended meanings.  Thus suppose that the advocates, Alice and Bob, of such a pair of dual theories are `wise enough' to use disjoint vocabularies. Nevertheless, their theories might well, once formalized, be definitionally equivalent. Indeed, for a duality in the  sense of our Schema, their theories no doubt {\em will} be definitionally equivalent, if they could have been rendered logically equivalent, by the argument of Section \ref{implic}, had Alice and Bob used the same vocabulary. Thus to repeat the Remark: despite the definitional equivalence, Alice and Bob are still making different  claims about the world! 

Essentially the same point comes up in the literature on theoretical equivalence, regardless of dualities. It is folklore among logicians that often, formal theories that they (anyone!) would think of as very different are definitionally equivalent: in effect, because the definitions added to one theory of the predicates etc. of the other are allowed to be `whatever it takes'  to deduce the theorems of the other theory. That is: the definitions are not required to be `intuitive' or to `respect meanings'. Hudetz (2016, Example 2; 2018, Proposition 4) gives an example which is striking, especially in the light of our previous discussion. He points out that a formal theory of Minkowski spacetime (postulating the affine space of $\mathR^4$, with a Minkowski interval) is definitionally equivalent to a formal theory of Euclidean geometry for  $\mathR^4$. The reason is that they are each a definitional extension of the theory of the real line $\mathR$. Roughly speaking: one `just' uses set theory to form appropriate $n$-tuples and to define an inner product with the desired signature. Then each can be definitionally extended to `encompass' the other. From within Minkowski geometry, one defines a second inner product with completely positive signature $(+,+,+,+)$; and vice versa, from within Euclidean geometry.

This is striking in two ways. (i): It illustrates the general idea that any two definitional extensions of a given theory that introduce different terms are definitionally equivalent since they can each add the other's definitions to yield a common definitional extension. (ii): Recall Section \ref{credo}'s example of two formulations of relativity theory with trivially swapped conventions about  $+/-$ in the metric's signature (used to emphasise that dual theories might well agree utterly in their claims).	Here, on the other hand, we naturally think of definite vs. indefinite signature as encoding important geometric differences: differences that the formal notion of definitional equivalence is blind to. Thus, taking the two examples together, we have a striking illustration of how interpretation moulds verdicts of theoretical equivalence.\\

Finally, I note that my Remark and Implication still hold good, beyond the case of definitional equivalence. To be specific:  Here are two other  weakenings of logical equivalence that have been proposed, and that have evident merits as explications of theoretical equivalence:---\\
\indent \indent \indent (A): {\em Generalized definitional equivalence}. This generalizes, i.e. weakens, the idea of definitional equivalence to deal with many-sorted logic: (Barrett and Halvorson (2016, 2016a), who call it  `Morita equivalence'). This generalization seems right, in that it enables us to say, for example, that a formulation of Euclidean geometry based on points, i.e. with points in the domains of its models, is equivalent to (`the same theory as') a formulation  based on lines: (Barrett and Halvorson (2017)). \\
\indent \indent \indent (B): {\em Categorical equivalence}, i.e. equivalence of categories in the sense of category theory. Note that this  notion is defined for {\em any} two categories. So the broad idea here is to argue that a scientific theory is best conceived as a category: the models of the theory (in some sense of `model'!) are the objects of the category, and the arrows of the category are appropriate maps between models. This is not the place for details about this broad idea: for which, cf. Halvorson and Tsementzis (2015), Weatherall (2015). I shall just make three remarks.\\
\indent \indent \indent \indent (i): The broad idea is motivated by the difficulty, emphasised in (B) of Section \ref{implic}, of capturing physical theories adequately and perspicuously in formal languages, especially with a first-order logic. For discussion, cf. e.g. Weatherall (2016: 1077, 1080: discussions of his Criteria 1 and 1').  \\
\indent \indent \indent \indent (ii): On the syntactic conception of theory, the idea is take the objects to be models in the sense of model theory, i.e. as in Section \ref{implic}; and the arrows to be elementary embeddings. It turns out that on this version, (A) implies (B) but not conversely: (cf. Barrett and Halvorson 2016a, Sections 5 and 6, especially theorems 5.6, 5.7). \\
\indent \indent \indent \indent (iii): But the idea has also been applied to the semantic conception, i.e. to physical theories formulated informally, as having a set of solutions. It applies very naturally to spacetime theories, for which a solution is an entire spacetime equipped with a dynamically possible history of matter and-or radiation fields: e.g.  vacuum electromagnetism in Minkowski spacetime. Indeed, each such solution is often called a `model'. The set of solutions/models is then taken as the objects of the category, and the arrows of the category are structure-preserving maps between models. This framework has the merit of articulating in a precise way such topics as when one theory has `surplus  structure' compared with another. But this is not the place for details about this line of thought. \\

What matters for this paper is whether the Remark's Implication still holds. As discussed: it is bound to hold for a notion of theoretical equivalence that is weaker than logical equivalence---which definitional equivalence is; and so is (A); and so is (B), when (B) is implemented on the syntactic conception of theory, as above. And the Implication still holds for the now-familiar reason:  the definitions of theoretical equivalence,  by (A) and by (B) on the syntactic conception, are formal---they    set aside intended meanings. \\

To sum up this Section: the Implication is worth emphasising just because: \\
\indent \indent  (i):  the drift of the literature on  `theoretical equivalence'  is to weaken the notion from the elementary conception of logical equivalence; while \\
\indent \indent  (ii): the Implication of the Remark is that  in some cases---indeed cases of dual theories---{\em logical equivalence is too weak}. Its verdict `these theories are equivalent' misses the fact that the theories (as interpreted!) disagree with each other.

 \section{Examples in string theory}\label{string}
  \subsection{The merits of advanced examples}\label{needex}
So far in this paper, I have established the Remark and its Implication by using  elementary examples of our logically weak Schema of duality (Section \ref{weak;exples}). This situation prompts two related questions, indeed doubts, one could  have about what we have achieved.

(Q1):  Using elementary examples prompts the question: Are there dualities in advanced physics that exemplify (Contr) or (Diff)---and might these duals, also, get formalized so as to be logically equivalent? If so, the Remark and Implication are upheld. We again conclude:  logical equivalence and its weakenings are too weak as explications of theoretical equivalence. But if not, one might suspect that the Remark and Implication are of limited interest, an artefact of using elementary examples. And the suspicion is the stronger because, as I have admitted,  bosonization is {\em not} an example of (Contr) or (Diff).  

(Q2):  The Schema being logically weak prompts the question: Could we---should we---strengthen the Schema, in such a way that cases of (Contr) and (Diff) are ruled out? Here, the questioner must admit that adding to our Schema the requirements (i)-(iii) which we set aside at the start of Section \ref{weak;exples}---that the duality isomorphism be non-obvious, logically strong and scientifically important---will {\em not} rule out all cases of (Contr) and (Diff). For  even if you think these requirements rule out Examples (1) and (2): no one could deny that Kramers-Wannier duality, and the other statistical mechanical dualities to which it leads, i.e. Example (3) of (Diff), satisfy these requirements.  But you might speculate that there are other ways to strengthen the Schema, so as to rule out cases of (Contr) and (Diff). 

My answers to both questions is, in short: \\
\indent \indent (Q1): There are dualities in (very!) advanced physics, viz. string theory, that can be taken as cases of (Contr), and perhaps there are also cases of (Diff): and for both sorts of case, the duals might get formalized so as to be logically equivalent. Besides: \\
\indent \indent (Q2): These dualities amply satisfy the requirements (i)-(iii), of being non-obvious, logically strong and scientifically important.\footnote{Admitted: you might yet speculate that there are other ways to strengthen the Schema, so as to rule out cases of (Contr) and (Diff).  But why search for them, instead of accepting that dualities can exemplify (Contr) and (Diff)?}  \\
In Sections \ref{Ex6} and \ref{Ex7}, I justify my answer to (Q1) by briefly discussing gauge/gravity duality and T-duality, respectively. (Even a brief discussion will make it clear that these dualities are non-obvious, logically strong and scientifically important---so that my answer to (Q2) needs no further justification.) \\

But before we start, I must admit to an important qualification.  String theory is of course a vast intellectual area. It is also speculative, not established, physics: both in the sense that it lacks empirical confirmation, and in the sense that its conceptual framework and technical formalism are not yet settled. Similarly for string dualities. This is a vast subject, and most dualities are conjectural: though conjectured to be exact, not approximate---and to be provable in a formalism that is, one hopes, over the next hill. This situation suggests two criticisms of any project that tries to classify a string duality in a philosopher's  neat pigeonholes, such as my (Contr) or (Diff):---\\
\indent \indent  (i): It makes the project  look premature. For in most cases, no one knows if in fact the duality concerned holds. And even if it is known to hold in some specific case or cases (given by judicious choices of spacetime dimensions and-or topology, and-or field content, and-or subset of states and quantities), the sceptic may say that this specific case is  not of scientific interest---returning us to (Q1) above.\footnote{Here is another collision of jargon: such a `case' or `set of judicious choices' is usually called a `model', as in `sigma-model', `XY-model', or indeed `Ising model'. More important: string theory can boast a few such cases of rigorous dualities: cf. footnote \ref{stringrigour}.} \\
\indent  \indent  (ii): It makes the project look naive: that is, naively realist about the string theories we now have, or can reasonably hope soon to have. For, as is often said, string theorists regard each of the string theories we now have as some sort of approximation to a theory, dubbed `M-theory', that we have yet to formulate: of which, they hope, our present theories will be various regimes, or sectors. From this perspective, the main scientific interest of  the dualities between our present string theories lies in the hints they give about how to formulate M-theory. (In fact, T-duality was a factor prompting the idea of M-theory: Witten (1995).) This returns us to the {\em heuristic function} of dualities, which this paper has set aside, especially since the warning, {\em Beware}, and footnote \ref{heuristic} of  Section \ref{Ex1}. And as mentioned there: it is to be expected that M-theory---the theory `behind the duals', as against their common core or bare theory---will render the dualities between our present string theories approximate. Thus it can seem that my project of classifying a string duality as to (Contr) or (Diff) `takes the duality too seriously'. 
  
To these charges, I reply: `Guilty---but unrepentant!' It is obvious from  my endorsement of intensional semantics (Section \ref{intsemics}), and  of construing subject-matters in terms of partitions on a set of possible worlds, or physical states (Section \ref{s-matters}), that my overall outlook in logic and ontology is robustly realist. And I aspire to be a scientific realist, drawing my world-picture, albeit fallibly, from our mature scientific theories.    Agreed: these overall outlooks are contentious---but I cannot defend them here. And agreed:  for theories  that are not established, and not even in their final formulation, the results drawn from interpreting them in this robustly realist manner, can only be tentative and fallible. But, say I: nonetheless worth stating.\footnote{For a more nuanced rationale for interpreting theories  that are not established, I recommend Huggett and W\"{u}thrich (2013: Section 3).} So in this tentative and fallibilist spirit, I proceed ...


 \subsection{(6): Gauge/gravity duality}\label{Ex6}
 `Gauge/gravity duality' is the umbrella term for dualities between a string theory (hence including a description of gravity) on a $D$-dimensional spacetime (the `bulk') and a quantum field theory (a gauge theory, with no description of gravity) on a  $(D - 1)$-dimensional space or spacetime that forms the bulk's boundary. The original, best understood, and best confirmed (though still conjectural) example is between a type IIB string theory on a ten-dimensional spacetime, with five dimensional anti-de Sitter (`AdS') spacetime as a factor, and a certain conformal field theory (`CFT': viz. ${\cal N} = 4$ super-Yang Mills theory)  on its boundary. So this example is called `AdS/CFT': an acronym which is also used for the general idea.
 
Obviously, I cannot here survey, even cursorily, this cluster of ideas. Let me just say:\\
\indent \indent (1): About the desiderata that a duality should be non-obvious and scientifically important:--- To glimpse the non-obviousness, note that the duality maps ($d_s, d_q$ of Section \ref{details}: the {\em dictionary} in physicists' jargon) pair some very disparate items. For example: the bulk's radial spatial coordinate (i.e. the coordinate orthogonal to the boundary) gets mapped to the energy scale of the boundary theory. (There are also some notions that get paired with themselves, i.e. `the dictionary respects synonymy': which, surrounded by such disparate pairings, is itself striking. For example, temperature in the bulk gets mapped to temperature in the boundary.) The duality is also useful: problems that are difficult, because strongly coupled, in the boundary theory are mapped into easier, weakly coupled, problems in the bulk. \\
\indent \indent (2): For introductions from a physics perspective, cf. e.g. Horowitz and Polchinski (2006), Polchinski (2017, Section 4), De Haro et al. (2016). There are several recent philosophical discussions. We can distinguish four topics. \\
\indent \indent \indent (i):  Rickles (2013), Teh (2013),  Dieks et al. (2015) and De Haro (2017) focus on the suggestion prompted by gauge/gravity duality, that gravity or spacetime is emergent. \\
\indent \indent \indent (ii): De Haro and others have analysed the relation between duality and gauge symmetry, especially in connection with gauge/gravity duality (De Haro et al. (2016)): which depends on a subtle analysis (De Haro (2017a)) of boundary conditions, about the conditions for a symmetry on one side of the duality to be `visible' from the other side.\\
\indent \indent \indent (iii): Closer to this paper's concerns, and using its Schema for duality: De Haro (2016: Section 2.1; 2016a Section 2.2) has addressed the question what is the common core i.e. bare theory, of which the bulk and boundary theories are models (i.e. models in our representation-theory sense from Sections \ref{idea} and \ref{details})---especially as regards spacetime concepts. In short, he argues that in the bare theory, the spacetime is the $(D - 1)$-dimensional boundary manifold equipped---not with a metric, but merely---with an equivalence class of them, under local conformal transformations.\\
\indent \indent \indent (iv): Finally, there is precisely this paper's concern: Do the duals agree? In De Haro's jargon: are they physically equivalent? As I reported at the end of Section \ref{others}, several authors say `Yes'. (For some of the quotations I gave there are about gauge/gravity duality, at least in part.)  \\

The {\em denouement} is clear. I  say `No'; or more modestly: `Saying `No' is tenable'. My reason is straightforward. Indeed, it returns us to my first example of the Remark, Example (1) in Section \ref{Ex1}. 

Imagine we had a rigorous proof of some specific gauge/gravity duality. Suppose  the bulk theory (the `left dual') says spacetime is five-dimensional ($D$ = 5); so the boundary theory, the right dual, says it is is four-dimensional. But both theories are putative `theories of everything', `toy cosmologies'. They are both about a single topic, the cosmos; in philosophers' jargon, the actual world. So the theories make contrary assertions about that single topic, the actual world, namely about the dimension of its spacetime. So this is a case of (Contr).  In terms of the simple logic---or rhetoric!---of the situation, we have come full circle, back to Example (1), with its contrary specifications of absolute rest in Newtonian gravitation; and with noteworthy Features 1) and 2)---a `toy cosmology' with a complete set of states in each dual.\\

Five points in clarification and support of this verdict: (1), (2) and (5) are the most important. (5) returns us to the contrast between the heuristic function of dualities, and the present project of interpreting dualities as now formulated.\\
\indent \indent (1):  Of course, saying that the duals are both about the cosmos,  the actual world, does not commit me to either of them being actually true. For nor does Section \ref{Ex1}'s verdict that Example (1) illustrates (Contr) commit me to Newtonian gravitation, with some or other specification of absolute rest, being actually true. I mean `about' the cosmos in the sense of `referring' to it (having it as `the target system', in an ugly current jargon); not `describing it accurately'.\\
\indent \indent (2): Agreed: there is an important contrast between Example (1) and this Example. Example (1) had a single  `knob' whose `setting' specified an absolute rest, and so one of the duals, without anything else being affected. So it was easy to take the duals as rival `toy cosmologies' about the single actual world. But here, the  two descriptions are so different---a $D$-dimensional world with gravity vs. a $(D - 1)$-dimensional one without gravity, but with conformal invariance---that one naturally says they are about different subject-matters or topics. (After all, the literature never calls the conformal field theories, and other quantum field theories that are boundary theories, `cosmologies': they are  theories of certain non-gravitational interactions between quantum particles.)  But obviously, this is just a matter of what one means by `subject-matter' or `topic': and this usage says nothing against Section \ref{s-matters}'s other (equally natural) usage of `subject-matter' as a taxonomy, i.e. a partition of a set of possibilities/ possible worlds.  Cf. also (5) below.   \\
\indent \indent  (3): I agree that my `So the theories make contrary...' implicitly assumes that the cosmos has only one spacetime, with a unique dimension---it cannot be both 5 and 4. This is, I take it, part of what is meant by each of the bulk and boundary theories being a putative cosmology, so that there is no question of either dual countenancing that the system (`sector of reality') it describes is one of a number of such systems /sectors. Note the contrast with our Examples (2) and (3) of (Diff): for position-momentum duality, there can be two quantum particles on distinct (non-intersecting) lines, and for Kramers-Wannier duality, there can be two Ising lattices.\\
\indent \indent (4): Agreed: one could redefine `cosmos' so as to allow a cosmos to comprise two or more mutually disconnected spacetime manifolds, whose dimensions could differ. In which case, one such cosmos might include a five-dimensional manifold accurately described by our duality's bulk theory, and a four-dimensional manifold accurately described by its boundary theory. But this would simply turn the duals'  disagreement into a case of (Diff), like Examples (2) and (3), rather than (Contr). \\
\indent \indent (5): I agree that, faced with the proven duality, there is a temptation or impulse to `quotient'. That is: either \\
\indent \indent \indent (i) to formulate the duals' common core, the bare theory (if we have no formulation or a defective or unperspicuous one), and-or \\
\indent \indent \indent (ii) to formulate another theory `behind the duals', of which they are approximations, not representations. \\
In short, to say: `the real truth lies in what is in common, or what is behind, the two duals'. This is indeed the heuristic function of dualities, first discussed in the warning, {\em Beware} of  Section \ref{Ex1}: and noted for string theory at the end of Section \ref{needex}. Recall the analogous temptation for Example (1): either \\
\indent \indent \indent (i) to move to Galilean (neoNewtonian) spacetime, or \\
\indent \indent \indent (ii) to move to geometrized gravity, such as in general relativity. \\
So I of course agree that this temptation is worthy: scientifically, heuristically, valuable---and accordingly stressed by physicists' discussions. But we must not let this temptation, oriented to the further development of our theories, distort the activity of interpreting the theories {\em as now formulated}. We can put this in terms of De Haro's notions of internal and external interpretations, as follows. Having the imagined proof of the specific gauge/gravity duality does not {\em ipso facto} give us an internal interpretation, since the proof will use the current external interpretations, which as De Haro (2016a, Section 2.2) says `are of course completely different in the two formulations of the theory: [for example] on the gravity side, the [boundary] manifold $M$ is the boundary of a $D$-dimensional manifold with a metric which can vary; whereas $M$ is a manifold in the QFT with a fixed metric.' Thus a statement of the bare theory, and an internal interpretation, are {\em not} automatic, given a proven duality. They are not just over the next hill. And so the duals as now formulated give a case of (Contr): just as Newton and Clarke would have said in Example (1).

  \subsection{(7): T-duality}\label{Ex7}
  `T-duality' is the umbrella term for two dualities between two pairs of string theories (as currently formulated): one pair called `type II' and the other `heterotic'. Both dualities involve inverting the radius of one of the compact dimensions of space: (for string theories require several more dimensions than four, so that each theory takes the extra dimensions to be compact, `curled up' like a circle). Thus a type IIA theory postulating that a certain dimension of space has radius $R$ is dual to a type IIB theory where the dimension is $1/R$. (Here, my `1' suppresses a constant.) Similarly for heterotic theories; but we can focus on just one case, say the type II case.
  
  There is an obvious objection to this idea---which has been addressed. The objection is that if one theory, say a type IIA theory, postulates a radius $R$ so small that it could not be empirically detected, $1/R$ may well be so large that it {\em could} be detected, i.e. could be detected if it was real. And since, `looking around us', we do not see detectably large compact dimensions, we should infer that the dual type IIB theory is disconfirmed, while the type IIA theory is not. So surely the duality is broken: the dual theories  are not empirically equivalent, let alone fully theoretically equivalent in the way that duals are meant to be.  
  
However, there is a reply to this objection: or at least, a consensus among string theorists how to reply! Again I cannot here survey, even cursorily, this large topic. It must suffice to say that if one analyses an archetypal measurement of the radius of a putative compact dimension---say by sending off a particle of known velocity, e.g. a photon, and timing how long it takes to return from its journey---one can argue that the result can be naturally accommodated by {\em both} dual theories. This is essentially because what one dual describes as a journey through physical space, the space described by familiar physics (and everyday thought and language), is  described by the other dual as a journey through an internal space (called `winding space'). This kind of analysis goes back to Brandenberger and Vafa (1989, Section 2); cf. also \'{A}lvarez et al. (1995: Section 6). For philosophers, a masterly exposition  is Huggett (2017: 83, and 84: `interpretation one'), whom I have followed.

Thus I shall take it that this objection can be laid to rest; so that in T-duality, the dual theories  are indeed empirically equivalent in a very strong sense---`theoretically equivalent', as they say! That is: the duals agree about the values of `all' quantities in `all' states, in the manner of eq \ref{obv1} (Section \ref{details}). So I can now face my main question. Do the duals agree in my sense: or do they disagree, either as (Contr) or as (Diff)? In De Haro's and Huggett's jargon: are they physically equivalent?\footnote{{\label{stringrigour}}{Besides, my question is less premature and-or naively realist than the accusations at the end of Section \ref{needex} might suggest. For thanks to the fact that T-duality does not exchange weak and strong coupling (though it does exchange large and small spatial radius), it can be proven order-by-order in string perturbation theory. This means that, broadly speaking, its status as `established' or `proven' is no worse than any perturbative results in a quantum field theory. Besides, there are rigorous proofs, within topological string theory, of {\em mirror symmetry}: of which T-duality is a special case, but for a richer theory. My thanks to Sebastian De Haro for these points.}} 

This question seems to have attracted less attention from philosophers than the parallel question about gauge/gravity duality, in Section \ref{Ex6}. But at least two authors say: `they agree'. Namely, Rickles (2011: 61;  2013a, 61; though cf. his footnote 14) and, in a much more detailed analysis,  Huggett (2017: 85-87, Section 2.2). Indeed, we glimpsed Huggett's views, in (c) at the end of Section \ref{others}, and in (2)(i) of Section \ref{symmthy}. Thus he first notes, in connection with harmonic oscillator duality, that the two duals are physically {\em inequivalent} (in my jargon: disagree)  because we can independently measure e.g. the mass by means that go beyond the oscillator, or our model of it. Then he writes: 
\begin{quote}
But the case is disanalogous to string theory, if that is taken as a theory of {\em everything} ... consider a world in which the harmonic oscillator is the complete physical description, so that there is no more encompassing theory by which mass or the spatial amplitude of oscillation could be uniquely determined. ... do we now say ... the parameters such as mass that differentiate the duals ... distinguish two distinct physical possibilities, which nevertheless agree on the values of all observables?

[He concedes that it] would not be a logical fallacy, nor [contravene] unavoidable semantic or ontological principles, [to deny that] the duals describe the same physical possibility. [But ...] from a practical scientific point of view, it makes sense to treat those differences as non-physical ... long established well-motivated scientific reasoning should lead us to think that dual total theories represent the same physical situation (2017: 86).
\end{quote}

Obviously, I beg to differ: along lines closely parallel to those in the {\em denouement} of Section \ref{Ex6}. That is: I interpret the duals as both being about a single topic: the cosmos, the actual world. They make contrary assertions about this topic. So they disagree: a case of (Contr). Besides, Section \ref{Ex6}'s five points of clarification and support carry over with trivial changes of wording: disagreement over the value of a spatial radius replaces disagreement over the number of spatial dimensions. In particular, point (5)'s diagnosis, that the literature misses my Remark, because of its focus on the heuristic function of dualities, carries over. But to avoid repeating myself, I just invite the reader to check that the five points carry over.

Finally, I should make two supplementary remarks. The first is about Huggett's views; the second about distinguishing types of string theory---which will take us back to Section \ref{Ex6}'s (3) and (4). \\
\indent \indent (i): In fairness to Huggett, I should report that after he says the duals agree, he goes on to address the pressing question that results: how can we make sense of the `appearance' that the dual theories contradict each other about the radius of space? He distinguishes two answers, called `interpretation one' (p. 84) and `interpretation two' (p. 85). In short: the first says that (some of) the words  must be given different interpretations on the two sides: which interpretations, and for which words, being judiciously chosen so as to ÔcompensateÕ for the apparent contradiction, i.e. chosen so that the duals in fact agree. So the appearance that the dual theories contradict each other is a mistake due to homonymy. On the other hand, the second says that each word is given the same interpretation on the two sides; and then it saves Huggett's initial claim, that the duals are physically equivalent, by maintaining that the duals' contradiction is about {\em unphysical} topics. Huggett ends (p. 87) by favouring the second, his `interpretation two'. My reply to all this is that I find Huggett's discussion admirably sensible: but of course, I do not need to answer his pressing question---to choose between his interpretation one and interpretation two. For my simple verdict, (Contr), avoids the question.  \\
\indent \indent (ii): So far, I have suppressed the fact that T-duality holds between different string theories, e.g. a type IIA and a type IIB. (As does Huggett.) If these are treated as theories  of everything (TOEs, `toy cosmologies'), as string theories usually are, then they are incompatible; and so we have a case of (Contr)---parallel to the discussion in Section \ref{Ex6}'s point (3). On the other hand, I agree that maybe one could treat such theories,	not as TOEs, but as both true in a single cosmos/possible world with, say a 10-dimensional space: namely by having the type IIA describe one compact dimension as radius $R$, and  the type IIB describe another compact dimension as radius $1/R$. This would simply turn the duals'  disagreement into a case of (Diff) rather than (Contr)---just like in Section \ref{Ex6}'s point (4). \\ \\

{\bf Acknowledgements}:---  This paper is dedicated to Graeme Segal: what an inspiration he is, to heart as much as to mind. I am very grateful to Sebastian De Haro. This paper owes a lot to his ideas (and patient explanations!)---about all Sections, but especially of course Section \ref{string}. For comments and correspondence, I am grateful to Graeme Segal; to the vertices of the Cambridge Simplex, here anonymized {\em \`{a} la} Bourbaki; and to T. Barrett, S. Blundell, E. Chen, R. Dawid, N. Huggett, L. Hudetz, A. Meehan, O. Pooley, J. Read, S. Rivat, B. Roberts, T. Teitel and C. W\"{u}thrich. I also thank audiences in Cambridge, Dubrovnik, New York and Oxford; and the editors, not least for their patience.  
 
\section{References}
Abraham, R. and Marsden, J.~(1978). {\em Foundations of Mechanics}, American Mathematical Society, Chelsea Publishing, vol. 364.\\
\\
\'{A}lvarez, E., \'{A}lvarez-Gaum\'{e}, L. and Lozano Y. (1995). An introduction to T-duality in string theory. {\em Nuclear Physics B: Proceedings} {\bf 41}, pp. 1-20. arxiv: hep-th/9410237.\\ 
\\
Andreka, H., Madarasz, J., Nemeti, I. and Szekely G.  (2012). A logic road from special relativity to general relativity. {\em Synthese}, {\bf 186}, 633-649.\\
\\
Andreka, H. and Nemeti, I. (2014). Definability Theory: Course Notes. {\em Department of Logic, ELTE, Budapest}. \\
\\
Barrett, T.  (2015). On the Structure of Classical Mechanics.  {\it British Journal for the Philosophy of Science}, {\bf 66}, 801-828. http://philsci-archive.pitt.edu/9603/ \\
\\
Barrett, T. (2017). Equivalent and Inequivalent Formulations of Classical Mechanics. http://philsci-archive.pitt.edu/13092/\\
\\
Barrett, T.~and Halvorson, H.~(2016). Glymour and Quine on Theoretical Equivalence. {\it Journal of Philosophical Logic}, 45(5), pp.~467-483.\\
 http://philsci-archive.pitt.edu/id/eprint/11341\\
\\
Barrett, T.~and Halvorson, H.~(2016a). Morita Equivalence. {\it The Review of Symbolic Logic}, 9(3), pp.~556-582 \\
http://philsci-archive.pitt.edu/id/eprint/11511\\
\\
Barrett, T.~and Halvorson, H.~(2017). Fron geometry to conceptual relativity. {\it Erkenntnis}, {\bf 82}, 1043-1063. http://
http://philsci-archive.pitt.edu/12605/ \\
\\
Black, R. (2000), Against Quidditism, {\em Australasian Journal of Philosophy}, {\bf 78}, 87-104.\\
\\
Brandenberger, R. and Vafa C. (1989). Superstrings in the early universe. {\em Nuclear Physics B} {\bf 316}, pp. 391-410. \\
\\
Brush (1967), History of the Lenz-Ising model, {\em Reviews of Modern Physics}, {\bf 39}, 883.\\
\\
Butterfield, J.~(2004). Between Laws and Models: Some Philosophical Morals of Lagrangian Mechanics. arxiv: physics/0409030; http://philsci-archive.pitt.edu/1937/ \\
\\
Butterfield, J.~(2006).  On Symmetries and Conserved Quantities in Classical Mechanics, in W. Demopoulos and I. Pitowsky (eds.), {\em Physical Theory and its Interpretation}, Springer: 43 - 99; Available at:   
   http://arxiv.org/abs/physics/0507192  and at http://philsci-archive.pitt.edu/2362/ \\
   \\
Butterfield, J.~and De Haro, S.~(2018). A Schema for Duality: Examples. In preparation.\\
\\
Button, T. and Walsh S.~(2018). {\em Philosophy and Model Theory}, Forthcoming OUP.\\
\\
Coffey, K.~(2014). Theoretical equivalence as interpretative equivalence. {\it British Journal for the Philosophy of Science}, {\bf 65}, 821-844.\\
\\
Dawid, R. (2013). {\em String Theory and the Scientific Method}. Cambridge: Cambridge University Press\\
\\
Dawid, R. (2017). String dualities and empirical equivalence. {\em Studies in History and Philosophy of Modern Physics}, {\bf 59}, pp.~21-29.\\
\\
de Bouvere, K. (1965). Synonymous theories. In {\em Symposium on the theory of models}. North Holland 
publishing company.\\
\\
Callender, C. and Cohen, J. (2006). There is no special problem about scientific representation. {\em Theoria} {\bf 21}, pp.  67-85; http://philsci-archive.pitt.edu/2177/ \\ 
\\
De Haro, S.~(2016). Spacetime and Physical Equivalence. Forthcoming in {\it Space and Time after Quantum Gravity}, Huggett, N. and W\"uthrich, C.~(Eds.), http://philsci-archive.pitt.edu/13243/\\
\\
De Haro, S.~(2016a). Duality and Physical Equivalence. \\
 http://philsci-archive.pitt.edu/id/eprint/12279 (This is an expansion of (2016); and the title has changed to `Spacetime and Physical Equivalence'). \\
\\
De Haro, S. (2017). Dualities and emergent gravity: Gauge/gravity duality.
{\em Studies in History and Philosophy of Modern Physics}, {\bf 59}, pp.~109-125.\\ 
\\
De Haro, S. (2017a). Invisibility of Diffeomorphisms. {\it Foundations of Physics}, {\bf 47}, pp.~1464-1497.\\
\\
De Haro, S. (2018). The heuristic function of dualities. {\em Synthese}; https://doi.org/10.1007/s11229-018-1708-9\\
\\
De Haro, S.,  Butterfield, J.N.~(2017). A Schema for Duality, Illustrated by Bosonization. Forthcoming in {\em Foundations of Mathematics and Physics one century after Hilbert}. Editor: Joseph Kouneiher. Collection Mathematical Physics, Springer 2017. arxiv: 1707.06681; http://philsci-archive.pitt.edu/13229/\\
\\
De Haro, S.,  Butterfield, J.N.~(2018). A Schema for Duality: Examples. In preparation.\\
\\
De Haro, S., Mayerson, D., Butterfield, J.N. (2016). Conceptual Aspects of Gauge/Gravity Duality, {\it Foundations of Physics}, 46 (11), pp.1381-1425. doi:~10.1007/s10701-016-0037-4. arxiv: 1509.0923; http://philsci-archive.pitt.edu/12296/\\
\\
De Haro, S., Teh, N., Butterfield, J.N.~(2017). Comparing Dualities and Gauge Symmetries. {\em Studies in History and Philosophy of Modern Physics}, {\bf 59}, pp.~68-80. arxiv: 1603.08334; http://philsci-archive.pitt.edu/12009\\
\\
Dieks, D., Dongen, J. van, Haro, S. de~(2015), Emergence in Holographic Scenarios for Gravity. 
{\it Studies in History and Philosophy of Modern Physics} {\bf 52}, pp.~203-216. doi:~10.1016/j.shpsb.2015.07.007.\\
\\
Goldblatt, R.~(2004). {\em Orthogonality and Spacetime Geometry}. Springer.\\
\\
Goldstein, H., Poole, C. and Safko, J. (2002). {\em Classical Mechanics}. Addison Wesley: third edition.\\
\\
Halvorson, H. (2012). What scientific theories could not be. {\em Philosophy of Science} {\bf 79}, pp.~183-206.\\
\\
Halvorson, H. and Tsementzis, D. (2015). Categories of scientific theories. Forthcoming in {\em Categories for the Working Philosopher}, edited by E. Landry. http://philsci-archive.pitt.edu/11923/\\
\\
Hawking, S. and Ellis, G.~(1973). {\em The Large-scale Structure of Spacetime}, Cambridge: Cambridge University Press.\\
\\
Hodges, W.~(1997). {\em A Shorter Model Theory}, Cambridge University Press.\\\
  \\
  Horowitz, G. and Polchinski, J. (2006). Gauge/gravity duality. In D. Oriti (ed.), {\em Towards Quantum Gravity?}, Cambridge: Cambridge University Press. arxiv: gr-qc/0602037\\
  \\
Hudetz, L. (2016). Definable categorical equivalence: towards an adequate criterion of theoretical intertanslatability. Manuscript, October 2016.\\
  \\
Hudetz, L. (2017). The semantic view of theories and higher-order languages. http://philsci-archive.pitt.edu/14084/\\
\\
Hudetz, L. (2018). Definable categorical equivalence. http://philsci-archive.pitt.edu/14297/\\
   \\
   Huggett, N. (2017). Target space $\neq$ space. {\em Studies in History and Philosophy of Modern Physics}, {\bf 59}, pp.~81-88. http://philsci-archive.pitt.edu/11638/ \\
   \\
  Huggett, N. and W\"{u}thrich, C. (2013). Emergent spacetime and empirical (in)coherence.  {\em Studies in History and Philosophy of Modern Physics}, {\bf 44}, pp.~276-285.\\
   \\
Jordan, T. (1969). {\em Linear Operators for Quantum Mechanics}, John Wiley. Dover reprint 2006.\\
\\
Lavis, D. and Bell, G.~(1999). {\em Statistical Mechanics of Lattice Systems 1: closed form and exact solutions}, Springer.\\
\\
Lewis, D. (1970). General semantics. {\em Synthese} {\bf 22}, pp.~18-67; reprinted in his {\em Philosophical Papers: volume 1}  (1983), Oxford University Press.\\
\\
Lewis, D. (1974). Radical Interpretation. {\em Synthese} {\bf 23}: 331Ð344: reprinted with Postscripts in his {\em Philosophical Papers: Volume I} (1983), Oxford: Oxford University Press.\\
\\
Lewis, D. (1975). Languages and Language. In Keith Gunderson (ed.), {\em Minnesota Studies in the Philosophy of Science, Volume VII}, Minneapolis: University of Minnesota Press: 3Ð35.\\
\\
Lewis, D. (1983). New Work For a Theory of Universals, {\em Australasian Journal of Philosophy} {\bf 61}: 343Ð377: reprinted in his {\em Papers in Metaphysics and Epistemology} (1999), Cambridge: Cambridge University Press.\\
\\
Lewis, D. (1983a). Putnam's paradox. {\em Australasian Journal of Philosophy} {\bf 62}: 221Ð236;  reprinted in his {\em Papers in Metaphysics and Epistemology} (1999), Cambridge: Cambridge University Press.\\
\\
Lewis, D. (1988), Relevant implication, {\em Theoria} {\bf 54}, 161-174; reprinted in his {\em Papers in Philosophical Logic}, Cambridge University Press 1998; page references to reprint. \\
\\
Lewis, D. (1988a), Statements partly about observation, {\em Philosophical Papers} {\bf 17},  1-31; reprinted in his {\em Papers in Philosophical Logic}, Cambridge University Press 1998; page reference to reprint. \\
\\
Lutz, S. (2017). What Was the Syntax-Semantics Debate in the Philosophy of Science About? {\em Philosophy and Phenomenological Research} {\bf 95}, 319-352. doi 10.1111/phpr.12221. http://philsci-archive.pitt.edu/11346/\\
\\
Malament, D. (2012). {\em Topics in the Foundations of General Relativity and Newtonian Gravitation Theory}. Chicago Lectures in Physics. Chicago: University of Chicago Press.\\
\\
Matsubara, K. (2013). Realism, underdetermination and string theory dualities. {\em Synthese} {\bf 190}, 471-489.\\
\\
M\"{o}ller-Nielsen, T. (2017). Invariance, interpretation and motivation. {\em Philosophy of Science} {\bf 84}, 1253-1264.\\
\\
Polchinski, J. (2017). Dualities of fields and strings. {\em Studies in the History and Philosophy of Modern Physics}, {\bf59}, pp.~6-20. doi:~10.1016/j.shpsb.2015.08.011.\\
\\
Pooley, O. (2019). {\em The Reality of Spacetime}. Forthcoming.\\
\\
Prugovecki, E. (1981). {\em Quantum Mechanics in Hilbert Space}, Academic; second edition. Dover reprint 2006.\\
\\
Putnam, H. (1962). The Analytic and the Synthetic. In H. Feigl and G. Maxwell (eds.) {\em Minnesota Studies in the Philosophy of Science} vol. III, Minneapolis: University of Minnesota Press,  358-397.\\
\\
Putnam, H. (1981). {\em Reason, Truth and History}, Cambridge: Cambridge University Press.\\
\\
Quine, W. (1953). Two Dogmas of Empiricism. In his {\em From a Logical Point of View}, Cambridge MA: Harvard University Press.\\
\\
Read, J. (2016). The interpretation of string-theoretic dualities. {\em Foundations of Physics} {\bf 46}, 209-235. http://philsci-archive.pitt.edu/11205/ \\
\\
Read, J.  and M\"{o}ller-Nielsen, T. (2018). Motivating dualities. Forthcoming in {\em Synthese}: http://philsci-archive.pitt.edu/14663/ \\
\\
Rickles, D.~(2011). A philosopher looks at string dualities. {\it Studies in History and Philosophy of Modern Physics}, 42, pp.~54-67.\\
\\
Rickles, D.~(2013). AdS/CFT duality and the emergence of spacetime. {\it Studies in History and Philosophy of Modern Physics}, {\bf 44}, pp.~312-320.\\
\\
Rickles, D.~(2013a). Mirror symmetry and other miracles in superstring theory. {\it Foundations of Physics}, {\bf 43}, pp.~54-80.\\
\\
Rickles, D.~(2017). Dual theories: `same but different' or `different but same'? {\em Studies in the History and Philosophy of Modern Physics}, 59, pp.~62-67. doi:~10.1016/j.shpsb.2015.09.005.\\
\\
Rynasiewicz, R. (1994). The Lessons of the Hole Argument, {\em The British Journal for the Philosophy of Science} {\bf 45}, 407-436. \\
\\
Savit, R. (1980), Duality in field theory and statistical systems, {\em Reviews of Modern Physics}, {\bf 52}, 453-487.\\
\\
Shoemaker, S. (1982). The inverted spectrum. {\em The Journal of Philosophy} {\bf 79}, pp. 357-381.\\
\\
Stein, H. (1992). Was Carnap Entirely Wrong, After All? {\em Synthese} {\bf 93}, 275--295.\\
\\
Takhtajan, L. (2008). {\em Quantum Mechanics for Mathematicians}, American Mathematical Society.\\
\\
Teh, N. (2013). Holography and emergence. {\it Studies in History and Philosophy of Modern Physics}, {\bf 44}, pp.~300-311.\\
\\
Teh, N. and Tsementzis, D. (2017). Theoretical equivalence in classical mechanics and its relationship to duality. {\it Studies in History and Philosophy of Modern Physics}, {\bf 59}, pp.~44-54.\\
\\
Thompson, C. (1972).  {\em Mathematical Statistical Mechanics}, Princeton: Princeton University Press.\\
\\
van Fraassen, B. (1989). {\em Laws and Symmetry}. Oxford: Oxford University Press.\\
\\
van Fraassen, B. (2014). One or two gentle remarks about Hans Halvorson's critique of the semantic view. {\em Philosophy of Science} {\bf 81}, pp. 276-283.\\
\\ 
Visser, A. (2006), Categories of Theories and Interpretations. In {\em Logic in Tehran. Proceedings of the
workshop and conference on Logic, Algebra and Arithmetic, held October 18Ð22, 2003.} {\em Association of Symbolic Logic}. \\
\\
Weatherall, J. (2015).  Categories and foundations of classical fields. Forthcoming in {\em Categories for the Working Philosopher}, edited by E. Landry. http://philsci-archive.pitt.edu/11587/; arxiv: 1505.07084.\\
\\
Weatherall, J. (2016). Are Newtonian  gravitation and  geometrized Newtonian  gravitation theoretically equivalent? {\em Erkenntnis}, {\bf 81}, pp. 1073-1091: http://philsci-archive.pitt.edu/11575/ \\
\\
Witten, E. (1995). String theory dynamics in various dimensions. {\em Nuclear Physics B}, {\bf 43}, 85. arxiv: hep-th/9503124 .

\end{document}